\documentclass[twocolumn,showpacs,reprint,amsmath,amssymb,aps,prb]{revtex4-1}
\usepackage{graphicx}
\usepackage{dcolumn}
\usepackage{bm}
\usepackage{subfigure}
\usepackage{natbib}
\usepackage{multirow}
\usepackage{soul}
\usepackage[usenames,dvipsnames]{xcolor}
\usepackage{soul}
\usepackage{floatrow}
\usepackage{amsmath}
\usepackage{amsfonts}
\usepackage[hyperfootnotes=false]{hyperref}
\usepackage{epstopdf}
\hypersetup{
 colorlinks=true,
 citecolor=blue,
 linkcolor=blue,
 urlcolor=Blue}
 
\begin{document}
\title{First principle studies on  electronic and thermoelectric properties of Fe$_{2}$TiSn based multinary Heusler alloys}

\author{Mukesh K. Choudhary\textit{$^{1,2}$} H. Fjellv\r{a}g\textit{$^{3}$}  and P. Ravindran\textit{$^{1,2}$}}
\email{raviphy@cutn.ac.in}

\affiliation{$^{1}$Department of Physics, School of Basics and Applied Sciences, Central University of Tamil Nadu, Thiruvarur, India}
\affiliation{$^{2}$Simulation Center for Atomic and Nanoscale MATerials (SCANMAT) Central University of Tamil Nadu, Thiruvarur, India.}
\affiliation{$^{3}$Center for Materials Science and Nanotechnology, Department of Chemistry, University of Oslo, PO Box 1033, N0315, Norway}

\begin{abstract}
The alloys with 8/18/24 valence electron count (VEC) are promising candidates for efficient energy conversion and refrigeration applications at low as well as high temperatures. Recently Fe$-$ based Heusler alloys attracted  researchers due to their compelling electronic band structure i.e flat band along one direction of the Brillouin zone and highly dispersive bands along the other directions. Here we focus on the thermoelectric (TE) transport properties of isovalent/aliovalent substituted Fe$_{2}$TiSn systems those may be potential TE materials. The multinary substitution has been done in such a way that it preserves the 24 VEC and hence the semiconducting nature. The calculated total energies with VASP$-$PAW potential within density functional theory with PBE$-$GGA functional were used to determine the ground state properties such as equilibrium lattice parameters, bulk modulus etc. We have also investigated the structural, electronic, lattice dynamic and TE transport properties by using PBE$-$GGA and TB$-$mBJ exchange$-$correlation functional.
The full potential linearized augmented plane wave method as implemented in WIEN2k code was used to investigate electronic structure and TE transport properties with the PBE$-$GGA and TB$-$mBJ exchange potentials and Boltzmann transport theory.  The calculated single crystal elastic constants, phonon dispersion and phonon density of states confirm that these systems are mechanically and dynamically stable.  The TE transport properties is calculated by including the lattice part of thermal conductivity ($\kappa_{L}$) obtained  from two methods one from the calculated  elastic properties calculation ($\kappa^{elastic}_{L}$) and the other from phonon dispersion curve ($\kappa^{phonon}_{L}$). The strong phonon$-$phonon scattering by large mass difference/strain fluctuation of isovalent/aliovalent substitution at Ti/Sn sites of Fe$_{2}$TiSn  reduces the lattice thermal conductivity which results in high \textit{ZT} value of 0.81 at 900\,K for Fe$_{2}$Sc$_{0.25}$Ti$_{0.5}$Ta$_{0.25}$Al$_{0.5}$Bi$_{0.5}$. The comparative analysis of TE transport properties using the band structures calculated with the PBE$-$GGA and TB$-$mBJ functional shows that the \textit{ZT} value obtained from TB$-$mBJ scheme is found to be significantly higher than that based on PBE$-$GGA. The calculated  relatively low lattice thermal conductivity and high \textit{ZT} values suggest that isovalent/aliovalent substituted Fe$_{2}$TiSn are promising candidates for medium to high temperature waste heat recovery.
\end{abstract}
\maketitle

\section{Introduction}
Present world is facing a great challenge to control the global greenhouse gas emissions. The global energy sectors set a target of reaching the net zero emission by 2050 limiting global average temperature to 1.5$^{\circ}$C. However, for limiting the warming to 1.5$^{\circ}$C the world need to take major and immediate actions. Although it is very challenging in keep the global warming to 1.5$^{\circ}$C. With the development of new technologies and scientific understanding world has the ability to tackle the climate change. Renewable energy technologies play an important role to achieve the target by supplying 70$-$80\% of electricity to the global need by 2050. Also, by reducing energy demand and improving the efficiency by converting the waste heat into useful clean form of energy have significant potential to reduce greenhouse emissions. Thermoelectric (TE) materials are capable of converting waste heat from industrial plants, vehicles, human body and other heat emitting devices into clean form of energy. The advantages of using TE materials like solid state operation, highly reliable, low maintenance, long life cycle, environment friendly and containing no moving part attracted many researchers to use them for energy conversion and refrigeration such as aerospace~\cite{bennett1996status, el2003efficient}, remote power, thermal energy sensors~\cite{ihring2011surface, escriba2005complete}, biomedical, industrial or commercial products and military applications~\cite{riffat2003thermoelectrics, chen2012effect, luo2013low, meng2011performance} using Seebeck and Peltier effects ~\cite{elsheikh2014review}. The efficiency of the TE materials are calculated by TE figure$-$of$-$merit 
\begin{equation} \label{eqn1} 
\textit{ZT} = {S^{2}\sigma T}/{\kappa},
\end{equation}
 where $S$,  $\sigma$ and $T$ are Seebeck coefficient, electrical conductivity and absolute temperature
 and $\kappa$ is the total thermal conductivity consist of electronic part $\kappa_{e}$ and the lattice part $\kappa_{L}$.  For high TE conversion efficiency,  materials should have high $S$ as well as $\sigma$ and low $\kappa_{L}$ values. The $S$, $\sigma$ and $\kappa_{L}$ are interrelated to each other, for example, $S$ and $\sigma$ are inversely related to each other which make one difficult to increase the TE power factor (PF; defined as S$^{2}\sigma$/$\tau$) above a particular value.  Among the metals, semiconductors, and insulators, the semiconductors  are considered as  good choice for TE materials due to their high electrical conductivity and relatively high $S$. On the other hand, metals have high electrical conductivity however,  at the same time they also have low $S$ value. In contrast, insulators have high $S$ value, but they lack high $\sigma$ value. These result very small TE conversion efficiency in metals as well as insulators and hence they are not considered as suitable materials for TE applications.
\par
With the discovery of Heusler alloys (HA) back in 1903 by Fritz Heusler opened the huge possibilities to design the endless number of compounds with the vast variety of properties ranging from spintronics ~\cite{balke2008rational, galanakis2007doping, zhang2018first}, optoelectronics~\cite{kacimi2014ii}, spin gapless semiconductors~\cite{ouardi2013realization, yang2016first, nelson2015enhancement}, ferromagnetism~\cite{hamaya2009ferromagnetism,krenke2006ferromagnetism}, thermoelectrics~\cite{zhu2019discovery,nishino2011development}, superconductors~\cite{sprungmann2010evidence, kierstead1985coexistence} topological insulators~\cite{chadov2010tunable, al2010topological, xiao2010half} to shape memory alloys~\cite{manosa2010giant,sutou2004magnetic,takeuchi2003identification}. 
The search for the HA as a high efficiency TE materials have accelerated in the recent years. The HA show excellent properties such as band gap tunability, non$-$toxic eco$-$friendly materials with semiconducting and magnetic behavior. Also, the HA are suitable for low to high temperature applications. Full Heusler (FH) alloys are ternary intermetallic compounds and generally they show a huge number of magnetic properties whereas half  Heusler (HH) alloys have attracted interest in the TE energy generation and refrigeration. 
\par
One of the most advantage of using HA is that by applying simple VEC rule one can design HA with various electronic behavior such as metallic/half$-$metallic, semiconducting and insulating nature. 
The total spin magnetic moment $\textit{M}_{t}$ of the HH and FH alloys are related to the total number of valence electrons $\textit{Z}_{t}$ by the relation $\textit{M}_{t}=\textit{Z}_{t}-2N_{\downarrow}$ where $\textit{Z}_{t}$ is the total number of electrons given by the sum of spin$-$up and spin$-$down electrons, while the total magnetic moment is given by the difference. This is known as Slater$-$Pauling rule.
\begin{flushleft}
\begin{equation}\label{eqn2} 
\textit{Z}_{t}=N_{\uparrow}+N_{\downarrow},  \hspace{0.2cm}  \textit{M}_{t}=N_{\uparrow}-N_{\downarrow}  \hspace{0.2cm} {\rightarrow}\hspace{0.2cm} \textit{M}_{t}=\textit{Z}_{\uparrow}-2N_{\downarrow}.
\end{equation}
\end{flushleft}
In the case of HH alloys if 9 minority bands are fully occupied then the total spin magnetic moment is given by the relation 
 \begin{equation}\label{eqn3} 
  \textit{M}_{t}=\textit{Z}_{t}-18.
 \end{equation}
For example, in HH alloy NiMnSb has 22 valence electrons which gives total moment of 4$\mu$B with half$-$metallic behavior and for TiNiSn it reduce to zero correspond to the non$-$magnetic and semiconducting behavior. In the case of  FH alloys, if 12 minority bands are fully occupied then the total spin magnetic moment is given by the relation  $\textit{M}_{t}=\textit{Z}_{t}-24$. For example, Co$_{2}$VAl and Fe$_{2}$VAl has 26 and 24 valence electrons show the total magnetic moment of 2 $\mu$B and non$-$magnetic (0 $\mu$B) with metallic and semiconducting behavior, respectively.  The HH alloys with 18 VEC show semiconducting behavior with high $S$ and power factor (PF) in low to high temperature range~\cite{choudhary2020thermal}. On the other hand, FH alloys mostly show metallic behavior leads to low PF and poor TE properties. However, FH alloys with 24 VEC exhibit semiconducting behavior with high PF. Similar to HH alloys FH alloys also has high thermal conductivity making the TE performance low. There has been lot of efforts to boost the TE performance of FH alloys by modulating their thermal conductivity.
\par
Recently Fe$_{2}-$based HA with general formula {Fe$_{2}$YZ} has gained high interest due to their very compelling structural, electronic and magnetic properties~\cite{gasi2013structural, ayuela1999structural}. The {Fe$_{2}$YZ} based alloys have been intensively studied for their potential applications in spintronics~\cite{kammerer2004co, sharma2013electronic}.  
Over the last few years search for the new TE materials based on FH has also been made~\cite{bilc2015low, sharma2014investigation}. The large PF in {Fe$_{2}$YZ} based FH alloys compared to the well$-$known classical TE materials (such as PbTe and Bi$_{2}$Te$_{3}$) implies that FH alloys are potential candidates for high efficiency thermoelectric applications~\cite{bilc2015low}. According to Slater$-$Pauling rule FH alloys with 24 VEC such as Fe$_{2}$VAl, Fe$_{2}$VGa, and Fe$_{2}$TiSn show semi$-$metallic or semiconducting properties~\cite{xu2008optical, lue2007thermoelectric, lue2008effects, slebarski2004electronic}. Both the experimental and theoretical studies show that the Fe$_{2}$VAl alloy is non$-$magnetic and exhibits pseudogap at the Fermi level~\cite{nishino1997semiconductorlike, bansil1999electronic, singh1998electronic, weinert1998hybridization, weht1998excitonic}. Recent studies show that both Fe$_{2}$TiSi and Fe$_{2}$TiSn have flat and dispersive bands in the conduction band i.e flat band along $\Gamma -$X direction and highly dispersive along other directions. 
\par
These alloys possess high Seebeck coefficient with the electron carrier concentration ranging from 1$\times$10$^{20}$ to 1$\times$10$^{21}$ cm$^{-3}$ at room temperature~\cite{yabuuchi2013large}.  It may be noted that materials with high density of states around conduction band minimum  (CBM) are suitable for the n$-$type TE materials. 
Daniel. Bilc \textit{et. al}~\cite{bilc2015low} proposed an approach for finding the high efficiency TE materials by achieving a narrow energy distribution around band edges and low carrier effective mass~\cite{mahan1996best} in bulk semiconductors without any nanostructuring or introduction of resonant states, and the theoretical concept is demonstrated in {Fe$_{2}$YZ} alloys. Another study showed that the flat bands with band width ~0.04 eV in Fe$_{2}$TiSn result in enhanced TE properties at room temperature~\cite{buffon2017thermoelectric, yabuuchi2013large}. Ilaria Pallecchi \textit{et al }~\cite{pallecchi2018thermoelectric} studied the effect of 10\% and 20\% Sb substitution at the Sn site in Fe$_{2}$TiSn using structural characterization, electrical, thermoelectrical, and thermal transport measurements. From these measurements they found that the 10\% Sb substitution at the Sn site in Fe$_{2}$TiSn increases the hole carrier concentration, inducing a weakly metallic behavior, while 20\% substitution lowers the carrier density and increases the resistivity by a factor of 50, restoring a semiconducting behavior, as in the undoped sample. Also, there are several studies on the influence of atomic disorder on the electronic structure and magnetic properties of  Fe$_{2}$TiSn ~\cite{slebarski2000weak,slebarski2006electron}. 
\par
In the present study, we have investigated the multinary substitution (both isovalent and aliovalent) on the FH alloys with VEC 24 and studied their electronic structure, lattice dynamics, chemical bonding and TE transport properties by using first principles theory. In the isovalent substitution case we have substituted Si at Sn site and Zr at Ti site of {Fe$_{2}$TiSn} and in another case we have substituted Ge at Sn site and Zr at Ti site of {Fe$_{2}$TiSn}. In the aliovalent substitution case we have substituted Sc/Ta at Ti site and Al/Bi at Sn site of {Fe$_{2}$TiSn}. For both the cases, we are able to preserve the 24 VEC and hence all the substituted systems are semiconducting in nature.  We have also calculated the electronic and TE transport properties by employing the TB$-$MBJ and PBE$-$GGA exchange$-$correlation  functional. 

\section{Computational Details}
Density functional theory (DFT) calculations for structural optimization and the electronic structure calculations were performed using projector$-$augmented plane$-$wave (PAW)~\cite{kresse1999ultrasoft} method, as implemented in the Vienna \textit{ab initio} simulation package (VASP)~\cite{kresse1996efficiency}. The generalized gradient approximation (PBE$-$GGA)~\cite{perdew1996generalized} proposed by Perdew$-$Burke$-$Ernzerh has been used for the exchange$-$correlation potential (V$_{xc}$) to compute the ground state parameters namely the lattice constant, the bulk modulus, the pressure derived bulk modulus and ground state energy. The irreducible part of the first  Brillouin zone (IBZ) was sampled using a Monkhorst pack scheme~\cite{monkhorst1976special} and employed a 12$\times$12$\times$12 (cubic systems) and 12$\times$12$\times$8 (tetragonal systems) \textbf{k}$-$mesh for geometry optimization. A plane$-$wave energy cutoff of 600 eV is used for geometry optimization for all the multinary substituted  Fe$_{2}$TiSn. The convergence criterion for energy was taken to be 10$^{-6}$ eV/cell for total energy minimization and for the convergence criterion for Hellmann$-$Feynman force acting on each atom was taken less than 1 meV/\AA~ to find the equilibrium positions. We have used the tetrahedron method with Bl\"{o}chl  correction~\cite{blochl1994improved} for IBZ integrations to calculate the density of states. Our previous study~\cite{choudhary2020thermal} show that the computational parameters used for the present study are sufficient enough to accurately predict the equilibrium structural parameters for multinary HH alloys. The full potential linearized augmented plane wave method as implemented in WIEN2k code~\cite{blaha2001wien2k,schwarz2003solid} was used to investigate electronic structure and TE transport properties with the PBE$-$GGA and Tran$-$Blaha modified Becke$-$Johnson  (TB$-$mBJ)~\cite{tran2009accurate} exchange potentials. We have used a very high density $\bf{k}$-$grid$ of \textbf{k}$-$point 34$\times$34$\times$34 for IBZ integration with R$_{MT}$K$_{max}$ $=$ 7 and the convergence criteria is set to be 1 mRy/cell for all our WIEN2k calculations in order to obtain accurate eigenvalues as well as transport properties. 
We have used the calculated eigen values and eigen vectors with high k$-$point density in the BoltZTraP code~\cite{madsen2006boltZTrap} for calculating the TE transport properties.  For calculating the lattice dynamic properties, a finite displacement method implemented in the VASP$-$phonopy~\cite{togo2015first} interface was used with supercell approach. We have used relaxed primitive cells to create supercell of dimension 2$\times$2$\times$2 with the displacement distance of 0.01\AA for phonon calculations.

\begin{figure*}
\centering
\includegraphics[height=8cm]{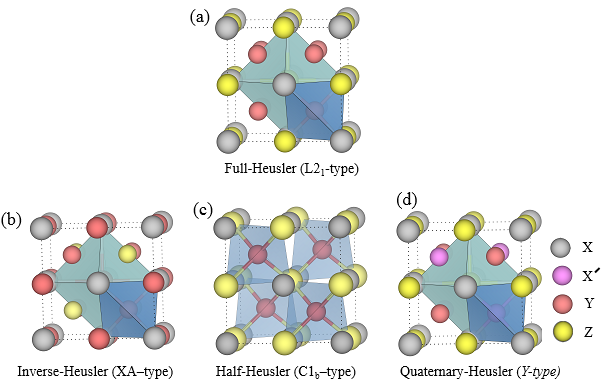}      
\caption{Different types of Heusler structures (a) full$-$Heusler , (b) inverse$-$Heusler, (c) half$-$ Heusler and, (d) quaternary$-$Heusler structures}
\label{fgr:cry}
\end{figure*}
\par 
In general, for periodic solids, one can calculate the electronic structure with the Kohn$-$Sham (KS)~\cite{kohn1965self} DFT~\cite{hohenberg1964inhomogeneous} by solving the following equation for the one$-$electron wave functions $\psi_{i, \eta}$

\begin{equation}\label{eqn4} 
\Big(-\frac{1}{2} \nabla^2 + \nu^{KS}_{eff,\eta} (r)\Big) \psi_{i,\eta} (r) = \epsilon _{i, \eta} \psi_{i,\eta} (r)
\end{equation}

In this equation $\nu^{KS}_{eff, \eta}$ = $\nu_{ext}$ + $\nu_H$ + $\nu_{xc,\eta}$ is the KS multiplicative effective potential (for spin $\eta$), which is the sum of the external ($\nu_{ext}$), Hatree ($\nu_H$), and exchange$-$correlation ($\nu_H$) terms, respectively. 
\par
One of the most challenging task of DFT in the KS formalism~\cite{kohn1965self} is to solve the band gap problem. The band gap from local density approximation (LDA)~\cite{perdew1992accurate} or generalized gradient approximation (GGA)~\cite{perdew1996phys} underestimate the corresponding experimentally measured band gap value of  semiconductors and insulators. To improve the band gap value of LDA/GGA comparable to the experimental values two methods have been mostly adopted. 
\par
The first method is hybrid functional approach where the LDA/GGA is mixed with an exact Fock exchange. The second method is screen exchange approach, in this approach the LDA/GGA correlation is combined with a screen non$-$local exchange and these approaches are implemented in the generealized KS equation~\cite{seidl1996generalized}. The effective potential used in these approaches are non$-$local, in contrast to that, the effective potential used in the standard KS equation is local potential. Another method widely used to overcome the bandgap problem is GW~\cite{aulbur2000exact,faleev2004all,van2006adequacy,chantis2007quasiparticle,shishkin2007self,shishkin2007accurate} method which can give very accurate band gap values compared to that from LDA/GGA functionals but it requires very expensive calculations.
The LDA and GGA approximations are the standard choice for calculating the exchange$-$correlation term. However, the LDA and GGA functionals are very good in predicting the equilibrium structural parameters and electronic structure of solids. But these functionals failed to predict the correct band gap value, which is too small compared to that obtained from experimental studies. The prediction of small band gap value is due to the self$-$interaction error in the LDA and GGA exchange$-$correlation potentials~\cite{perdew1981density}.  
\par
In order to obtain better band gap values with an accuracy comparable to that from experiments Becke and Johnson (BJ)~\cite{becke2006simple} proposed a very simple approximation which is totally density dependent and does not require two$-$electron integration for the exact exchange optimized effective potential (OEP),~\cite{sharp1953variational,talman1976optimized} that was similar to the Talman$-$Shadwick~\cite{talman1976optimized}  potential in atoms and also it is computationally less expensive compared to other methods used to predict band gap values accurately. 
Later in 2009 Trans and Blaha (TB)~\cite{tran2009accurate} found that the BJ exchange potential is still underestimate the band gap and proposed a simple modification in the original BJ potential which is given as 

\begin{equation}\label{eqn5} 
\nu^ {MBJ}_{ x,\eta} (r) = c \nu ^{BR} _{x,\eta} (r)  + ( 3c - 2) \frac{1}{\pi}  \sqrt{\frac{5}{12}} \sqrt{\frac{2t_{\eta} (r)}{\rho_{\eta} (r)}},
\end{equation}

where $\rho_\eta$  = $\sum^{N_\eta} _{i=1}$ $\left| \psi_{i,\eta}\right| ^{2}$ is the electron density, t$_{\eta}$ = $\frac{1}{2}$   $\sum^{N_\eta} _{i=1}$ $\nabla$ $\psi^{*}_{i,\eta}$ $\cdot$  $\nabla$ $\psi_{i,\eta}$ is the kinetic $-$ energy density, and 

\begin{equation}\label{eqn6} 
\nu^ {BR}_{ x,\eta} (r) = - \frac{1}{b_{\eta } (r)} \Big( 1 - e^{-x_{\eta} (r)} -  \frac{1}{2} x_{\eta} (r)  e^{-x_{\eta (r)} }\Big ),
\end{equation}

is the Becke$-$Roussel (BR) ~\cite{becke1989exchange} potential which was proposed to model the Coulomb potential created by the exchange hole.  Originally, BJ used the Slater potential $\nu^{Slater}_{x,\eta}$ ~\cite{slater1951simplification} instead of $\nu^ {BR}_{ x,\eta}$, but they showed that these two potentials are quasi$-$identical for atoms~\cite{becke2006simple}. In Eq. (5), the $c$ was chosen such a way that it depend linearly on the square root of the average of  $|\nabla \rho|/\rho$ :

\begin{equation}\label{eqn7} 
c = \alpha + \beta \Big( \frac{1}{V_{cell}} \int_{cell} \frac{|\nabla_\rho (r')|}{\rho(r')} d^3 r' \Big)^{1/2},
\end{equation}

where $ \alpha$  and $\beta$ are two free parameters and V$_{cell}$ is the unit cell volume. The modified BJ potential known as mBJ potential results band gap values with more accuracy than the BJ potential. Also it is less computational demanding compared with hybrid and GW methods. Using the TB$-$mBJ potential the band gap values of many semiconductors/insulators are calculated accurately and are comparable to that from experimental studies~\cite{tran2009accurate,koller2011merits,koller2012improving,jiang2013band}.

\begin{table*}
\small\addtolength{\tabcolsep}{-2pt}
\caption{\ The equilibrium structural parameters for isovalent/aliovalent substitution at Ti and Sn sites in Fe$_{2}$TiSn where Z=1;  A,  B, C, D, E and F are in the Wyckoff  position 8g (x$^{'}$, y$^{'}$, z$^{'}$), 1a (0, 0, 0), 3c (0, 1/2, 1/2), 1b (1/2, 1/2, 1/2) and 3d (1/2, 0, 0) for cubic and  4i (1/2, 0, 1/4), 1a (0, 0, 0), 1b (0, 0, 1/2), 1c (1/2, 1/2, 0), 1d (1/2, 1/2, 1/2) and 8r ((x$^{'}$, y$^{'}$, z$^{'}$), 1a (0, 0, 0), 1b (0, 0, 1/2), 1c (1/2, 1/2 0), 1d (1/2, 1/2, 1/2),  2e (1/2, 0, 0)  for tetragonal systems respectively. The lattice parameters $ a $ as well as $c$ (in \AA)  and the internal structural parameters (x$^{'}$, y$^{'}$, z$^{'}$) are obtained from our structural optimization,  heat of formation ($\Delta$H$ _{f} $ ; in kJ mol$^{-1}$), equilibrium volume (\AA$^{-3}$), bulk modulus (B$_{0}$),  pressure derivative of bulk modulus (B$_{0}^{'}$), PBE$-$GGA (E$_{g}^{PBE-GGA}$) and TB$-$mBJ (E$_{g}^{TB-mBJ}$) band gap values (in eV) are also listed. All the multinary substituted cubic and tetragonal systems have space group Pm$\bar{3}$m (No. 221) and  P4/mmm (No. 123), respectively.}
\label{tbl:1}
\begin{tabular*}{\textwidth}{@{\extracolsep{\fill}}ccccccccccccccc}            
\hline
Compound&\multicolumn{2}{c}Unit$-$cell dimension (\AA) &&\multicolumn{3}{c} Positional parameters&$\Delta$H$ _{f}$ & V & (B$_{0}$)& (B$_{0}^{'}$)&E$_{g}^{PBE-GGA}$ & E$_{g}^{TB-mBJ}$\\ 
\cline{2-3} \cline{5-7} \\
&a&c&&x$^{'}$&y$^{'}$&z$^{'}$&(kJ mol$^{-1}$)&(\AA$^{-3}$)\\ 
 \hline \\ 
Fe$_{2}$TiSn &\hspace{0.2cm}6.04&&&&&&\hspace{-0.27cm}$-$30.0&220.67&\hspace{0.2cm}182.53& \hspace{0.2cm}4.15&0.02&0.61\\ [1.0 ex]
Fe$_{2}$Ti$_{0.75}$Zr$_{0.25}$Sn$_{0.75}$Si$_{0.25}$&\hspace{0.2cm}6.01&&&0.237&\hspace{0.2cm}0.237&0.237&\hspace{-0.27cm}$-$39.78&216.92&\hspace{0.2cm}188.98&\hspace{0.2cm}4.16&0.07&0.62\\ [1.0 ex]
Fe$_{2}$Ti$_{0.5}$Zr$_{0.5}$Sn$_{0.5}$Si$_{0.5}$&\hspace{0.2cm}4.22&6.02&&&&&\hspace{-0.27cm}$-$49.50&107.49&\hspace{0.2cm}191.34&\hspace{0.2cm}4.14&0.13&0.41\\ [1.0 ex]
Fe$_{2}$Ti$_{0.25}$Zr$_{0.75}$Sn$_{0.25}$Si$_{0.75}$&\hspace{0.2cm}5.95&&&0.261&\hspace{0.2cm}0.261&0.261&\hspace{-0.27cm}$-$53.56&210.93&\hspace{0.2cm}195.93&\hspace{0.2cm}4.35&0.22&0.74\\ [1.0 ex]
Fe$_{2}$Ti$_{0.75}$Zr$_{0.25}$Sn$_{0.75}$Ge$_{0.25}$&\hspace{0.2cm}6.03&&&0.240&\hspace{0.2cm}0.240&0.240&\hspace{-0.27cm}$-$34.75&219.28&\hspace{0.2cm}184.61&\hspace{0.2cm}4.56&0.03&0.59\\ [1.0 ex]
Fe$_{2}$Ti$_{0.5}$Zr$_{0.5}$Sn$_{0.5}$Ge$_{0.5}$&\hspace{0.2cm}4.25&6.06&&&&&\hspace{-0.27cm}$-$40.05&109.69&\hspace{0.2cm}183.50&\hspace{0.2cm}4.37&0.08&0.62\\ [1.0 ex]
Fe$_{2}$Ti$_{0.25}$Zr$_{0.75}$Sn$_{0.25}$Ge$_{0.75}$&\hspace{0.2cm}6.12&&&0.259&\hspace{0.2cm}0.259&0.259&\hspace{-0.27cm}$-$40.87&217.96&\hspace{0.2cm}186.42&\hspace{0.2cm}4.27&0.07&0.64\\ [1.0 ex]
Fe$_{2}$Sc$_{0.25}$Ti$_{0.5}$Ta$_{0.25}$Al$_{0.5}$Bi$_{0.5}$ & \hspace{0.2cm}6.08&6.02&&0.760&\hspace{0.2cm}0.760&0.244&\hspace{-0.29cm}$-$36.18&221.16&\hspace{0.2cm}174.16&\hspace{0.2cm}4.47&0.07&0.60\\ [1.0 ex]
 \hline
 \end{tabular*} 
 \end{table*}
\section{Results and Discussion}
\subsection{Structural Description }
\begin{figure}[h]
\centering
\includegraphics[height=8cm]{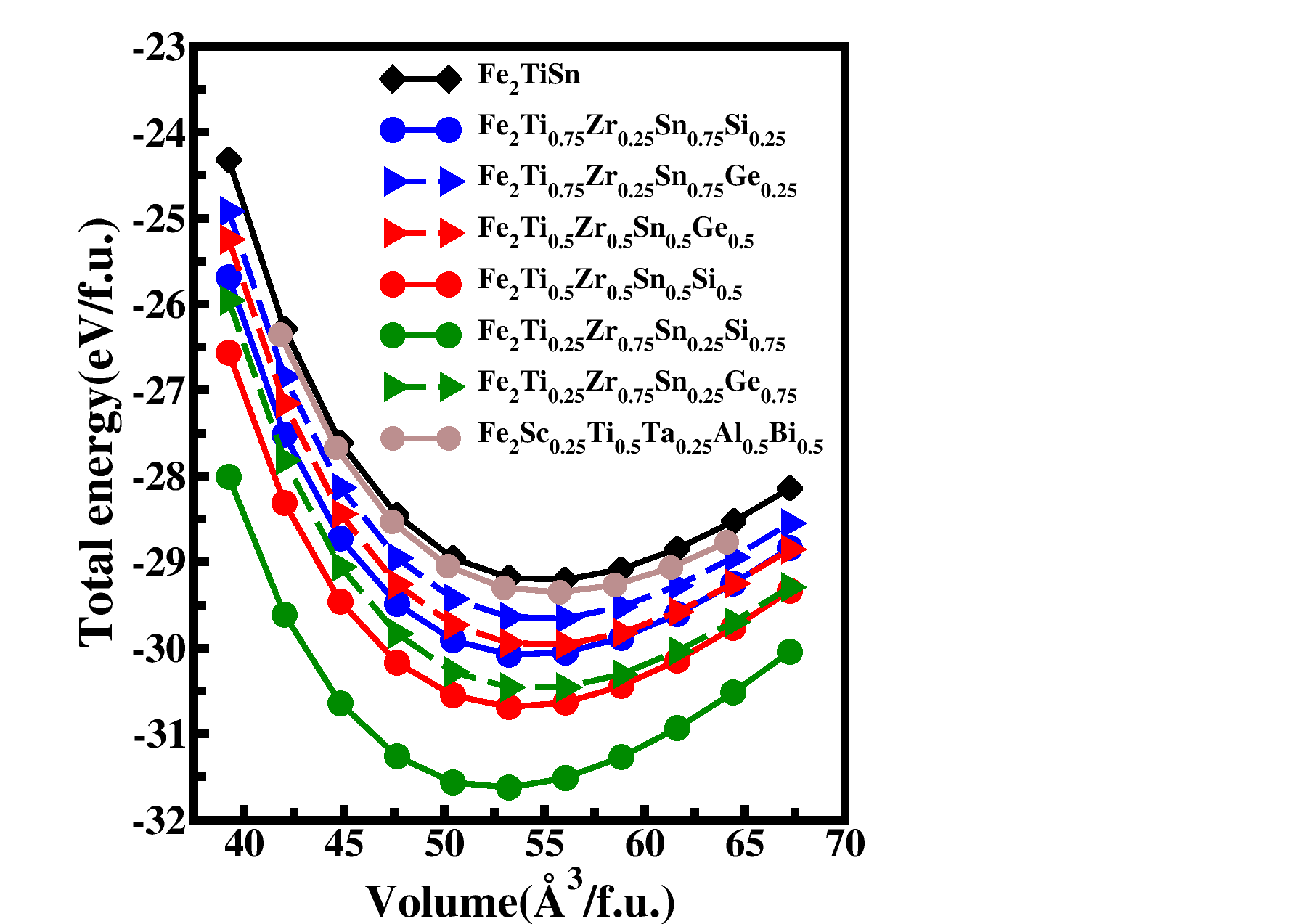}      
\caption{Calculated total energy vs volume curves for pure and isovalent/aliovalent  substituted Fe$_{2}$TiSn alloys obtained from PBE$-$GGA calculation.}
\label{fgr:1}
\end{figure}

Heusler alloys are divided into four main structural classes. A schematic representation for different types of  Heusler structure are presented in Fig \ref{fgr:cry}. First is FH alloys with the \textit{X$_{2}$YZ} composition. The FH alloys are possesing L$_{21}$ or Cu$_{2}$MnAl prototype (space group no 225: Fm$-$3m) crystal structure. Interestingly if \textit{X} atoms in the L$_{21}$ structure are replaced with a \textit{Y} or \textit{Z} atoms we will obtain a second family of HA  named as inverse HA  (\textit{XA}) having CuHg$_{2}$Ti  prototype (space group no 216: F$\bar{4}$3m) structure. Usually, the \textit{XA} structure is observed when the atoms at the \textit{Y} site has higher valence than that in the \textit{X} site. The cubic L$_{21}$ structure consists of four interpenetrating fcc sublattices~\cite{kandpal2007calculated} where the two \textit{X} atoms are equally placed on the Wyckoff position 8c (1/4, 1/4, 1/4). In contrast, \textit{Y} and \textit{Z} atoms are located at 4a (0, 0, 0) and 4b (1/2, 1/2, 1/2) positions, respectively. The third family of  HA is called half Heusler alloys, which has \textit{XYZ} composition and can be obtained by removing half of the \textit{X} atoms from FH alloys (\textit{X$_{2}$YZ} ). The HH alloys crystallize in the cubic structure (space group no 216 : F$\bar{4}$3m) with MgAgAs (C$_{1b}$) as a prototype. Finally, the fourth family of HA is obtained when one \textit{X} atom is replaced by a diferent atom \textit{X'} from the FH alloys in the L$_{21}$ structure. This is known as equiatomic quaternary Heusler alloys (EQHAs) with the \textit{XX'YZ} composition and LiMgPdSn as prototype with space group no 216: F$\bar{4}$3m. In this composition the\textit{ X} and \textit{X'} are different transition metals whereas the  \textit{Y } and \textit{Z} sites are occupied by a transition metal and a main group element, respectively. EQHA such as CoFeMnSi and inverse FH alloys (CuHg$_{2}$Ti) have been identified to be spin gapless semiconductors (SGS)~\cite{tsidilkovski2012electron} where one spin channel resembles that of a semiconductor, while the other has a zero band gap at the Fermi level and these materials gained high interest in tunable spin transport based applications~\cite{wang2008proposal,skaftouros2013search}.
Interestingly, electronic and magnetic properties of the Heusler family can be predicted by valence electron count (VEC) rule~\cite{tobola2000electronic,offernes2007electronic,kandpal2006covalent}. Previous studies show that Fe$_{2}$TiSn with 24 VEC is a non$-$magnetic semiconductor. To obtain the ground state properties we have carried out the volume optimization of  Fe$_{2}$TiSn, Fe$_{2}$Ti$_{1-x}$Zr$_{x}$Sn$_{1-x}$Si$_{x}$ and  Fe$_{2}$Ti$_{1-x}$Zr$_{x}$Sn$_{1-x}$Ge$_{x}$ where $x$=0, 0.25, 0.5 or 0.75, and Fe$_{2}$Sc$_{0.25}$Ti$_{0.5}$Ta$_{0.25}$Al$_{0.5}$Bi$_{0.5}$. The ground state energy as a function of equilibrium cell volume was computed with the PBE$-$GGA functional. The calculated relaxed lattice parameters, equilibrium volumes (\AA$^{-3}$), bulk modulus (B$_{0}$), and the pressure derivative of bulk modulus (B$_{0}^{'}$) are calculated from the total energy vs volume curve by fitting with Birch$-$Murnaghan's equation of state~\cite{murnaghan1944compressibility} 
  \begin{equation}\label{eqn9} 
  E_{tot}(V) = \frac{B_{0}V}{B^{'}_{0}(B^{'}_{0}-1)}\Big[\Big(\frac{V_0}{V}\Big)^{B^{'}_{0}}+B^{'}_{0} \Big(1-\frac{V_0}{V}\Big)-1\Big]+E_0
  \end{equation}
 where E is the energy, V$_{0}$ and V represent volume of the compound at zero pressure and finite pressure, respectively. B$_{0}$ and B$^{'}_{0}$ are isothermal bulk modulus and its pressure derivative at V=V$_{0}$. The total energy$-$versus$-$volume curves for these compounds are shown in Fig. \ref{fgr:1}. Table \ref{tbl:1} represents the calculated equilibrium lattice parameters computed by fitting with Birch$-$Murnaghan's equation of state, the heat of formation ($\Delta$H$ _{f} $), the calculated band gap value using PBE$-$GGA (E$_{g}^{PBE-GGA}$) and TB$-$mBJ$-$GGA ( E$_{g}^{TB-mBJ}$) exchange$-$ correlation functionals, bulk modulus (B$_{0}$), pressure derivative of  bulk modulus (B$_{0}^{'}$)  for  Fe$_{2}$TiSn, Fe$_{2}$Ti$_{1-x}$Zr$_{x}$Sn$_{1-x}$Si$_{x}$, Fe$_{2}$Ti$_{1-x}$Zr$_{x}$Sn$_{1-x}$Ge$_{x}$, and Fe$_{2}$Sc$_{0.25}$Ti$_{0.5}$Ta$_{0.25}$Al$_{0.5}$Bi$_{0.5}$. Our calculated equlibrium lattice constant for Fe$_{2}$TiSn~( 6.04 \AA) is in good agreement with the corresponding experimental value~\cite{slebarski2006electron} ~( 6.07 \AA).
  
 \begin{figure*}
\centering
\includegraphics[height=10cm]{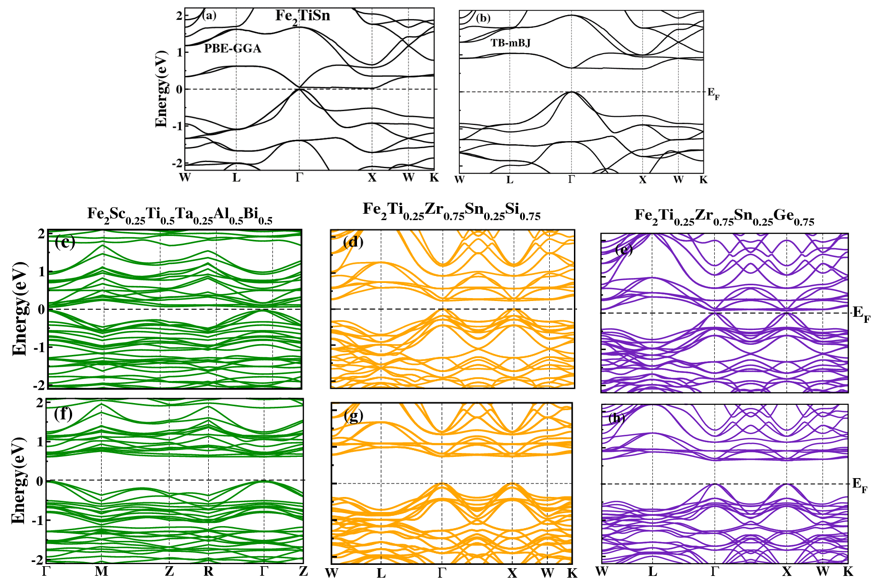}      
\caption{The calculated band structures for Fe$_{2}$TiSn, Fe$_{2}$Sc$_{0.25}$Ti$_{0.5}$Ta$_{0.25}$Al$_{0.5}$Bi$_{0.5}$, Fe$_{2}$Ti$_{0.25}$Zr$_{0.75}$Sn$_{0.25}$Si$_{0.75}$ and  Fe$_{2}$Ti$_{0.25}$Zr$_{0.75}$Sn$_{0.25}$Ge$_{0.75}$ obtained from PBE$-$GGA and TB$-$mBJ calculations are in (a,c,d,e) and (b,f,g,h), respectively. The Fermi level is represented by dashed line and is set to zero}.
\label{fgr:2}
\end{figure*}
\par
\subsection{Analysis of the electronic structure of multinary substituted Fe$_{2}$TiSn.}
It has been found that the band gap values calculated using TB$-$mBJ potential are in good agreement with the experimental value compared with that from LDA or GGA functionals ~\cite{koller2011merits}. Generally, the calculated band gap values using PBE$-$GGA is much smaller than that obtained from TB$-$mBJ potential. We have carried out electronic structure calculation by considering the high symmetry directions of the first IBZ of face centered cubic (when $x$=0, 0.25 and 0.75) and simple tetragonal (when $x$=0.5) lattice for the systems considered in the present study. The energy band gap values obtained using the PBE$-$GGA and TB$-$mBJ functionals are summarized in Table \ref{tbl:1}. In this section, we will describe the electronic band structure of pure and some of the selected isovalent/aliovalent substituted Fe$_{2}$TiSn systems and the band structures of other considered compounds are given in supplementary information. The electronic band structures calculated from the PBE$-$GGA functional show that these system possess a narrow band gap values ranging from 0.02 to 0.22\,eV depending upon the value of $x$. However, if we consider the TB$-$mBJ functional for our electronic structure calculations the calculated band gap values increases substantially compared with that from the PBE$-$GGA calculations and the calculated values vary from 0.41 to 0.74\,eV with $x$ as given in Table \ref{tbl:1}. The isovalent substituted systems with $x$=0.25 and 0.75 show direct band gap behaviour with valence band maximum (VBM)  and conduction band minimum (CBM) are at $\Gamma$ point. However, system with $x$=0.5 substitution  show indirect band gap behavior with VBM lying between M and Z and the CBM at $\Gamma$ point. Moreover, the aliovalent substitution with 25\% Sc/Ta at the Ti site and 50\% Al/Bi at the Sn site also show the direct band gap behavior with both VBM and CBM lying at $\Gamma$ point. The calculated band gap values clearly indicate that the results obtained with TB$-$mBJ are always larger than those obtained using PBE$-$GGA funtional. It may be noted that the band gap values obtained from PBE$-$GGA are always underestimate with those obtained from experimental values and hence one could expect that the band gap values obtained from TB$-$mBJ functional in the present study will be in good agreement with experiment. Unfortunately no experimental band gap value measurements are available for these systems to compare our results.
\par
Figure.\ref{fgr:2} shows the calculated band structures close to their band edges position i.e. $-$2 to 2 eV for Fe$_{2}$TiSn, Fe$_{2}$Sc$_{0.25}$Ti$_{0.5}$Ta$_{0.25}$Al$_{0.5}$Bi$_{0.5}$, Fe$_{2}$Ti$_{0.25}$Zr$_{0.75}$Sn$_{0.25}$Si$_{0.75}$ and  Fe$_{2}$Ti$_{0.25}$Zr$_{0.75}$Sn$_{0.25}$Ge$_{0.75}$ obtained using PBE$-$GGA and TB$-$mBJ functional. First, we will discuss the band structure of the pure Fe$_{2}$TiSn system. Figure. \ref{fgr:2} (a), (b) show the electronic structure of  Fe$_{2}$TiSn calculated from PBE$-$GGA and TB$-$mBJ functionals, respectively. The calculated band structures show the indirect band gap behavior where VBM is located at the $\Gamma$ point, and the CBM is along the X point. The bands located in the vicinity of  VBM are triply degenerate at the  $\Gamma$ point. However, the degeneracy is lifted when one move away from the $\Gamma$ point. Also, these degenerate bands creating well dispersed band around the VBM and hence one can expect hole conductivity in this system will be high. In the conduction band one can observe two different scenarios. Along  $\Gamma-X$ direction a band at the vicinity of CBM shows a very flat band dispersion and this flat band behaviour exists irrespective of the exchange correlation functional we have used as evident from Fig.\ref{fgr:2} (a) and (b). So, one can expect that the electron conductivity will be low in this direction due to high electron effective mass value. Hence, a large Seebeck coefficient is expected along this direction as the Seebeck coefficient is directly proportional to the effective mass. On the other hand, along $\Gamma-$L direction in the band structure just above the CBM we have noticed relatively well dispersed two degenerate bands. The similar band features have been observed in the case of  TB$-$mBJ computed band structure for Fe$_{2}$TiSn also but with the large band gap value compared with that from PBE$-$GGA functional calculation. The calculated indirect band gap values for Fe$_{2}$TiSn using PBE$-$GGA and TB$-$mBJ functionals are 0.025 and 0.61\,eV, respectively, consistent with the values from other reported works ~\cite{shastri2018comparative,meinert2013modified}. 
\par
Now let us discuss the band structures for the isovalent/aliovalent substituted Fe$_{2}$TiSn. In the case of isovalent/aliovalent substituted systems the calculated band gap values using GGA$-$PBE functional show that these systems possess narrow band gap values  (0.07$-$0.22 eV). However, the calculated band structure from TB$-$mBJ functional increases the band gap values of these systems as expected and the calculated values are in the range of 0.41$-$0.74 eV. Compared with two nearly degenerate bands just above CBM in Fe$_{2}$TiSn system present in the $\Gamma-X-W$ directions are well localized in isovalent/aliovalent substituted Fe$_{2}$TiSn systems as evident from Fig. \ref{fgr:2}. Such flat band behaviour at the CBM has strong influence on electron transport in these materials. Especially, these localized narrow bands at the CB edge will reduce the electronic part of thermal conductivity as well as electrical conductivity of these materials due to high electron effective mass which results in low electron mobility. In contrast, the two bands in the top most energy in the valence band not localized significantly by the  isovalent/aliovalent substitution and hence well dispersed  bands present at the VBM i.e. at $\Gamma$ point and hence the hole effective mass in these systems will be much lower than the electron effective mass. As a consequence of this one would expect that the hole conductivity in these systems will be higher than the conductivity from electrons. Though the electronic structure analysis suggest high power factor (S$^{2}\sigma$) in the doped system due to high  electrical conductivity by holes, the narrowing of  bands in the CB edge make enhancement in the Seebeck coefficient in the n$-$type doping condition by the presently attempted isovalent/aliovalent substitution and hence the power factor. So, one would expect increase in \textit{ZT} by n$-$type doping in isovalent/aliovalent substituted  Fe$_{2}$TiSn. 

\begin{table}
\centering
\caption{The carrier effective masses at the band edges (unit of free electron mass m$_{0}$) for Fe$_{2}$TiSn, Fe$_{2}$Ti$_{0.25}$Zr$_{0.75}$Sn$_{0.25}$Si$_{0.75}$, Fe$_{2}$Ti$_{0.25}$Zr$_{0.75}$Sn$_{0.25}$Ge$_{0.75}$ and Fe$_{2}$Sc$_{0.25}$Ti$_{0.5}$Ta$_{0.25}$Al$_{0.5}$Bi$_{0.5}$ calculated numerically by fitting the calculated electron dispersion curves from PBE$-$GGA and the TB$-$mBJ functional in the parabolic approximation}
\label{tbl:2}
\small\addtolength{\tabcolsep}{-1pt}
\begin{tabular}{cccccc} 
\hline \hline
\\[0.1ex]
Compound / direction & \multicolumn{2}{c}{PBE$-$GGA (eV)} & & \multicolumn{2}{c}{TB$-$mBJ (eV)}\\
\cline{2-3} \cline{5-6}
 &m$^{*}_{e}$ & m$^{*}_{h}$ &&  m$^{*}_{e}$ &m$^{*}_{h}$\\
\\ [0.01ex]
\hline
Fe$_{2}$TiSn &  &  &  &  & \\[0.5ex]
$\Gamma$ $-$ X & 52.19 & 0.72 &&  54.28 &0.74 \\
X $-$ $\Gamma$ & 38.99 & 6.85 &&37.69 &7.66 \\
X $-$ W& 0.84 & 0.69 && 1.08 & 2.21 \\
$\Gamma$ $-$ L & 0.33 & 0.36 &&   1.07 & 0.95 \\
Fe$_{2}$Ti$_{0.25}$Zr$_{0.75}$Sn$_{0.25}$Si$_{0.75}$ &  &  &  &  & \\[0.5ex]
$\Gamma$ $-$ X & 10.28 & 0.70 & &  13.70 & 0.64 \\
X $-$ $\Gamma$ & 10.28 & 0.70 & &  13.43 & 0.63 \\
X $-$ W & 9.76 & 0.69 && 14.13 & 0.64 \\
$\Gamma$ $-$ L & 0.84 & 0.45 && 1.60 & 0.76\\
Fe$_{2}$Ti$_{0.25}$Zr$_{0.75}$Sn$_{0.25}$Ge$_{0.75}$& &  &  &  & \\[0.5ex]
$\Gamma$ $-$ X & 17.90 & 0.59&& 20.87 & 0.62 \\
X $-$ $\Gamma$ & 17.90 & 0.64 && 23.34 &0.59 \\
X $-$ W &7.79 &0.61 & & 23.80 & 0.58 \\
$\Gamma$ $-$ L & 1.20 & 0.33 && 2.0& 0.74 \\
Fe$_{2}$Sc$_{0.25}$Ti$_{0.5}$Ta$_{0.25}$Al$_{0.5}$Bi$_{0.5}$&  &   & &  & \\[0.5ex]
$\Gamma$ $-$ Z &1.06 &0.60 && 1.22 & 0.53 \\
$\Gamma$ $-$ M & 1.30 & 0.533 & & 1.74 & 0.80 \\
$\Gamma$ $-$ R & 2.40 &0.75 & & 3.01 & 0.87 \\ \hline \hline
\end{tabular}
\end{table}

\par
Effective mass of charge carrier is considered as a key parameter for designing higher efficiency TE materials. The effective mass can be calculated from parabolic band approximation using an energy dispersion relation  
\begin{equation}\label{eqn8} 
\frac{1}{m^{*}_{b}} = \frac{1}{\hbar}\frac{\partial^2E}{\partial k^2},
\end{equation}
where $E$ , $\hbar$ ,$k$ and m$^{*}_{b}$ are the energy of electronic states, the reduced Planck's constant, the crystal momentum, and the band effective mass. As discussed in Sec.\ref{sub:Thermoelectric transport properties}, the m$^{*}_{d}$ in equation \ref{eqn26} refers to the density of states effective mass, which is given as m$^{*}_{d}$=N$_{v}$m$^{*}_{b}$ where N$_{v}$ is the band degeneracy and m$^{*}_{b}$ is the band effective mass. The Seebeck coefficient ($S$) can be increased by increasing the N$_{v}$ and m$^{*}_{b}$. However, a high value of m$^{*}_{b}$ results to low carrier mobility. Hence, increasing N$_{v}$ is considered as an effective way to increase the $S$ without affecting the carrier mobility.  It is well known that the electrical transport properties are directly related to quantities such as S, $\sigma$ and K$_{e}$ and can be tuned by tuning the band structure of the materials. In this study, the electron and hole effective masses of  pure and  isovalen/aliovalent substituted Fe$_{2}$TiSn were evaluated numerically by fitting the calculated dispersion curves from PBE$-$GGA and the TB$-$mBJ functional in the parabolic approximation. Based on the calculated band structure, we have computed the effective masses of electrons and holes for the pure as wells as isovalent/aliovalent substituted Fe$_{2}$TiSn. The effective mass is calculated along the high symmetry directions in the IBZ and the calculated effective masses are in the unit of free electron mass (m$_{0}$). In Table \ref{tbl:2} we have listed the calculated electron effective mass (m$^{*}_{e}$) and hole effective mass (m$^{*}_{h}$) at the CBM/VBM for the selected systems considered in the present study. The $\Gamma-X$ (i.e. (100) direction in real space) in the Table $\Gamma-X$ represents the effective mass calculated at $\Gamma$ point along the $\Gamma-X$ direction. The calculated  m$^{*}_{e}$ and m$^{*}_{h}$  for pure Fe$_{2}$TiSn  in high symmetric directions are given in Table \ref{tbl:2} and are agreeing well with the corresponding computed data reported in ref. [89]~\cite{shastri2018comparative}. However, there is no experimental effective mass measurements for any of the materials considered in the present study available to compare our results.
The hole/electron effective mass of the materials is inversely related to the curvature of the bands in the electronic band structure. This implies that the effective mass for the dispersed band is lower than that of the flat band. This concept is well satisfied with our calculated carrier effective mass values (see Table \ref{tbl:2}) for pure and isovalent/aliovalent substituted  Fe$_{2}$TiSn. From the Table \ref{tbl:2}, one can see that the calculated m$^{*}_{e}$ value for pure Fe$_{2}$TiSn at conduction band along the $\Gamma-$X direction is relatively higher than that along the $\Gamma-$L direction. This is due to the fact that the lowest energy conduction band along the $\Gamma-$X direction is more flat than that along the $\Gamma-$L direction. However, the bands around the VBM along the $\Gamma-$X  as well as $\Gamma-$L directions are more dispersive nature than the bands in the CB edge and hence the calculated value of m$^{*}_{h}$ the band in the VB edge is lower than the m$^{*}_{e}$ as given in Table \ref{tbl:2}.  From the Table. \ref{tbl:2} one can see that the isovalent/aliovalent substituted Fe$_{2}$TiSn systems show drastic decrease in the electron effective mass than that of pure Fe$_{2}$TiSn along the $\Gamma-X$ and $X-\Gamma$ directions irrespective of the exchange$-$correlation functional considered in the present study. On comparing the electron/hole effective mass of pure and isovalent/aliovalent substituted systems, we can see that the electron effective mass calculated from both PBE$-$GGA and TB$-$mBJ approaches show a higher value than that of hole effective mass. This is because the band in VB edge is more dispersed than that in the CB edge  as evident from Fig. \ref{fgr:2}.
\subsection{Mechanical stability and lattice dynamic calculation for pure and isovalent/aliovalent substituted Fe$_{2}$TiSn}\label{sub:Mechanical stability and lattice dynamic calculation for pure and isovalent/aliovalent substituted Fe$_{2}$TiSn}
The mechanical properties such as elastic constants, strength and fracture toughness are crucial parameters for designing the high efficiency TE devices for practical applications over a wide range of operating temperatures. The elastic response of a solid to different mechanical stress are described by elastic constants such as bulk modulus (B) (which measure the material's resistance against compression), shear modulus (G) (which can measure the response of the material to volume and shape change), Young modulus (E) (which measure the resistance against uniaxial tensions) and Poisson's ratio (v) (which is defined as the ratio of  the transverse contraction of a material to the longitudinal extension strain in the direction of the stretching force). The elastic modulus of  TE materials, for example, Bi$_{2}$Te$_{3}$, PbTe, SiGe, Skutterudites, and HH alloys~\cite{kallel2013thermoelectric, he2015studies, zhao2008thermoelectric} are very close to common engineering metals~\cite{davis1990metals} such as Al (70 GPa) and steels (200 GPa). It is found that the HH/FH alloys exhibits considerably higher hardness and modulus values and lower brittleness as compared with other TE materials. The material with high value of elastic moduli is suitable candidate to use in TE power generators, where both the mechanical stability and energy conversion efficiency are important. 
According to Hooke's law, the stress is proportional to the strain for small stress i.e. under elastic limit. The generalized form of stress$-$strain Hooke's law under the homogeneous deformation of crystal is given as  
$\eta_{ij}$  $=$ c$_{ijkl}$E$_{kl}$  where $\eta_{ij}$ and E$_{kl}$ are the homogeneous two$-$rank stress and strain tensors, respectively and c$_{ijkl}$ denotes the fourth$-$rank elastic stiffness tensor and can be described by a 6$\times$6 matrix (36 elements). Matrix representation of single crystal elastic constants in the Voigt notation is given as
\begin{equation}\label{eqn10} 
   \left({\begin{array}{cccccc}  \eta_{ij} \\
    \eta_{ij} \\
    \eta_{ij}\\
     \eta_{ij} \\
     \eta_{ij} \\
     \eta_{ij} \\
   \end{array}}\right)  = 
\left(\begin{array}{cccccc} 
c_{11} & c_{12} & c_{13} & c_{14} & c_{15} & c_{16} \\
c_{12} & c_{22} & c_{23} & c_{24} & c_{25} & c_{26} \\
c_{13} & c_{23} & c_{33} & c_{34} & c_{35} & c_{36} \\
c_{14} & c_{24} & c_{34} & c_{44} & c_{45} & c_{46} \\
c_{15} & c_{25} & c_{35} & c_{45} & c_{55} & c_{56} \\
c_{16} & c_{26} & c_{36} & c_{46} & c_{56} & c_{66} \\
\end{array}\right)
\left(\begin{array}{cccccc} 
   \epsilon_{ij} \\
   \epsilon_{ij} \\
   \epsilon_{ij} \\
   \epsilon_{ij} \\
   \epsilon_{ij} \\
   \epsilon_{ij} \\
 \end{array}\right), 
\end{equation}
where $\epsilon_{i}$, $\eta_{i}$ and c$_{ij}$ are strain, stress and single crystal elastic (stiffness) constants, respectively. The single crystal elastic constants can be calculated by applying the strain $\epsilon_{i}$ and calculating the corresponding stresses $\eta_{i}$  from the equation \ref{eqn10}.

 \begin{table*}[t]
 \centering
\caption{The calculated single crystal elastic constants c$_{ij}$ (in GPa), the bulk modulus (B in GPa), shear modulus (G in GPa), Young modulus (E  in GPa), longitudinal, transverse, average elastic wave velocity ($ \nu$\textsubscript{l} ,$\nu$ \textsubscript{t},$\nu$\textsubscript{m} in m/s), the Debye  temperature from elastic constants $\theta_{e}$ (in \,K), the Debye temperature from acoustic modes ($\theta_{a}$  in \,K) and Gr\"{u}neisen parameter ( $\gamma_{e}$)  based on changes in the stress by various strains within the elastic limit. Also, the calculated Debye temperature ($\theta_{\omega}$ in \,K) and the Gr\"{u}neisen parameter $\gamma_{\omega}$ using phonon dispersion curve based on finite difference method, primitive unit cell volume (V in (\AA $^{3})$), average mass per atom $\overline{M_{a}}$ (in amu), Poisson's ratio ($\nu$)  for Fe$_{2}$TiSn, Fe$_{2}$Ti$_{0.25}$Zr$_{0.75}$Sn$_{0.25}$Si$_{0.75}$, Fe$_{2}$Ti$_{0.25}$Zr$_{0.75}$Sn$_{0.25}$Ge$_{0.75}$, and Fe$_{2}$Sc$_{0.25}$Ti$_{0.5}$Ta$_{0.25}$Al$_{0.5}$Bi$_{0.5}$ obtained for the optimized structure with  PBE$-$GGA functional.}
\label{tbl:3}
\small\addtolength{\tabcolsep}{-0.1pt}
\begin{tabular}{ccccc}
\hline \hline
\\[0.1ex]
Parameters&Fe$_{2}$TiSn & Fe$_{2}$Ti$_{0.25}$Zr$_{0.75}$Sn$_{0.25}$Si$_{0.75}$ & Fe$_{2}$Ti$_{0.25}$Zr$_{0.75}$Sn$_{0.25}$Ge$_{0.75}$& Fe$_{2}$Sc$_{0.25}$Ti$_{0.5}$Ta$_{0.25}$Al$_{0.5}$Bi$_{0.5}$\\
\\ [0.5ex]
\hline
\\[0.05ex]
c\textsubscript{11} & 339.0& 375.30 & 291.90&318.99\\[0.8ex]
c\textsubscript{12} & 119.50 &121.90& 98.50&115.79\\[0.8ex]
c\textsubscript{13} & && &117.59\\[0.8ex]
c\textsubscript{44} &106.90 & 101.10 & 79.30&103.69\\[0.8ex]
c\textsubscript{33} & &  & &334.79\\[0.8ex]
c\textsubscript{66} & &  & &102.49\\[0.8ex]
B & 192.67& 206.37 & 162.97&186.03\\[0.8ex]
G& 108.03 & 110.67 & 85.86&103.43 \\[0.8ex]
E& 273.06& 281.66& 219.10&264.7\\[0.8ex]
B/G & 1.78&1.86& 1.89&1.80\\[0.8ex]
$ \nu$\textsubscript{l} & 7092.04& 6417.90 &9975.59&8733.67\\[0.8ex]
$\nu$ \textsubscript{t} & 4017.16 & 11476.98& 5549.44&4937.83\\[0.8ex]
$\nu$\textsubscript{m}&4467.10 &7144.24 &6180.06&5488.68\\[0.8ex]
V &55.17&211.18&230.06&221.15\\[0.8ex]
$\nu$ & 0.261 & 0.272& 0.275&0.261\\[0.8ex]
$\gamma_{e}$ & 1.54& 1.61 & 1.62&1.57\\[0.8ex]
$\gamma_{\omega}$&1.08 & 1.60 &1.55 &1.85\\[0.8ex]
$\overline{M_{a}}$ &74.13&68.34 &77.25&94.26\\[0.8ex]
$\theta_{e}$ &503.85&610.75&513.46&491.01\\[0.8ex]
$\theta_{a}$ &318.87&244.62 &205.65&196.66\\ [0.8ex]
$\theta_{\omega}$ & 277.52&200.48  &172.16 &164.96\\[0.8ex]\hline \hline
\end{tabular}
 \end{table*}

The number of independent single crystal elastic constants to describe the elastic properties of a crystal is depending on the symmetry of the crystal. The lower the symmetry, the more the number of independent elastic constants to describe the elastic properties. For example, the number of independent deformation matrices applied to cubic, hexagonal, trigonal, tetragonal, orthogonal, monoclinic, and triclinic crystals are 3, 5, 6, 6, 9, 13, and 21, respectively~\cite{levy20011,huntington1958elastic}. Single$-$crystal elastic constants can be obtained from first$-$principles calculations.~\cite{ravindran1998density} Several methods have been proposed for calculating the elastic constants of materials based in ab$-$initio total energy calculation. Most of the methods for calculating single crystal elastic constants are based on fitting the total energies or strain$-$stress relation of deformed crystals.~\cite{ravindran1998density,mayer2003ab,panda2006first} Many DFT codes such as VASP,~\cite{kresse1996efficient}, WIEN2k,~\cite{blaha2020wien2k} Quantum Espresso,~\cite{giannozzi2009quantum} CASTEP,~\cite{clark2005first} CRYSTAL,~\cite{perger2009ab,dovesi2018quantum} ABINIT,~\cite{gonze2002first,romero2020abinit} and SIESTA~\cite{soler2002siesta} could be employed to generate the elastic stiffness tensor of materials to remarkable accuracy. 
In the present study, elastic constants, c$_{ij}$ were extracted by applying a uniform deformation within the elastic limit to the relaxed crystal structure and calculated the yield stress as implemented in VASP code using PBE$-$GGA potential. 
Tables \ref{tbl:2} summarize the first$-$principle predicted elastic stiffness constants c$_{ij}$'s for pure and substituted isovalent/aliovalent substituted Fe$_{2}$TiSn systems at relaxed equilibrium volumes. The elastic properties of cubic and tetragonal phases can be fully described through 3 independent elastic constants such as (c\textsubscript{11}, c\textsubscript{12}, and c\textsubscript{44}) and 6 independent elastic constants such as (c\textsubscript{11}, c\textsubscript{12},c\textsubscript{13}, c\textsubscript{33}, c\textsubscript{44}, and c\textsubscript{66}), respectively. The other theoretical details of the calculation of mechanical properties from elastic constants, can be found elsewhere ~\cite{soderlind1993theory,ravindran1998density, jamal2014elastic,sneddon1958classical,huntington1958elastic}.
Based on the calculated c$_{ij}$, the mechanical stability for a given structure can be predicted according to Born stability criteria.~\cite{born_1940}  
\par
For cubic and tetragonal system this mechanical stability criteria are given below 
\begin{equation}\label{eqn11} 
c_{11}-|c_{12}|>0, \hspace{0.1cm}
 c_{11}+2c_{12}>0, \hspace{0.1cm}
  c_{44}>0
\end{equation}
\begin{equation}\label{eqntetra}
c_{11}>|c_{12} |, \hspace{0.1cm}2c_{13}<c_{33}(c_{11}+c_{12}), \hspace{0.1cm}c_{44}>0, c_{66}>0
\end{equation}

From the calculated c$_{ij}$ it is found that the investigated compounds are mechanically stable as they obey the stability criteria for the cubic and tetragonal structure. The polycrystalline elastic constants such as bulk modulus, shear modulus, Young's modulus, and Poisson's ratio can be calculated from the elastic stiffness moduli from the Voigt$-$Reuss$-$Hill (VRH) ~\citep{hill1952elastic} averaging approximations as
\begin{equation} \label{eqn12}
B_v = 1/9(c_{11}+c_{22}+c_{33})+2/9(c_{12}+c_{23}+c_{13})
\end{equation}
\begin{equation} \label{eqn13}
\hspace{-1cm}G_V=1/15(c_{11}+c_{22}+c_{33})-1/15(c_{12}+c_{23} +c_{13}
+1/5(c_{44}+c_{55}+c_{66}) 
\end{equation} 
 \begin{equation} \label{eqn14}
1/B_R = (s_{11} + s_{22} + s_{33}) + 2(s_{12} + s_{23} + s_{13}) 
\end{equation}
 \begin{equation} \label{eqn15}
    \hspace{-1cm}1/G_R =4/15 (s_{11} + s_{22} + s_{33}) -4/15 (s_{12} + s_{23} + s_{13}) 
     +1/5 (s_{44} + s_{55} + s_{66})
\end{equation}
here s$_{ij}$ is compliance constant i.e. the inverse matrix of c$_{ij}$. Finally the B and G are obtained by averaging the B$_{V}$ and  B$_{R}$, G$_{V}$ and G$_{R}$ as follows.
\begin{equation}\label{eqn16}
B=\frac{B_{V}+B_{R}}{2}
\end{equation}
\begin{equation}\label{eqn17}
G=\frac{G_{V}+G_{R}}{2}
\end{equation}
The Young's modulus (E) and the Poisson's ratio ($\nu$) were calculated using the following relations
\begin{equation} \label{eqn18}
E = \frac{9BG}{3B+G} 
\end{equation} 
\begin{equation} \label{eqn19}
\nu = \frac{3B-2G}{6B+2G} 
\end{equation} 
Furthermore the transverse (v$_{t}$) , longitudinal  (v$_{l}$) and average (v$_{m}$) sound velocities are calculated through the following equations  
\begin{equation} \label{eqn20}
v_t = \sqrt{\dfrac{E}{2\rho(1+\nu)}} = \sqrt{\frac{G}{\rho}}
\end{equation}  
\begin{equation} \label{eqn21}
v_{l}=\dfrac{E(1-\nu)}{\rho(1+\nu)(1-2\nu)} = \sqrt{(\frac{3B+4G}{3\rho})} 
\end{equation}  
\begin{equation} \label{eqn22}
v_{m}=\sqrt{[\frac{1}{3}(\frac{2}{v\textsubscript{t}\textsuperscript{3}}+\frac{1}{v\textsubscript{l}\textsuperscript{3}})]} 
\end{equation} 
Here $\rho$ is the mass density of the material.
From the above calculated v$_{t}$ , v$_{l}$ and v$_{m}$ one can calculate the Gr\"{u}neisen parameter ($\gamma$), Debye temperature ($\theta_{e}$) and acoustic Debye temperature ($\theta_{a}$) using the following equations 
  \begin{equation}\label{eqn23}
 \gamma=\dfrac{9-12(v\textsubscript{t}/v\textsubscript{l})^{2}}{2+4(v\textsubscript{t}/v\textsubscript{l})^{2}}
 \end{equation} 
 \begin{equation} \label{eqn24}
 \theta\textsubscript{e}=\frac{h}{k}[ \frac{3n}{4\pi}(\frac{N\textsubscript{A}\rho}{M})]\textsuperscript{1/3} v\textsubscript{m} 
 \end{equation} 
 \begin{equation} \label{eqn25}
 \theta\textsubscript{a}=\theta\textsubscript{e}n^{-1/3}
 \end{equation}
The calculated single crystal elastic constants, the bulk modulus, shear modulus, Young's modulus, and Poisson's ratio are listed in Table. \ref{tbl:3}. For the pure and isovalent/aliovalent substituted systems, our calculated bulk and shear modulus are varying in the range 162.97$-$206.37 and 85.86$-$186.03GPa, respectively. The difference between the elastic constants c$_{12}$ and c$_{44}$ (i.e. c$_{12}-$c$_{44}$) is known as Cauchy pressure, proposed by Pettifor~\cite{pettifor1992theoretical} which gives the information about the brittle/ductile behavior of a solid. A brittle material has negative Cauchy pressure, whereas a ductile material has positive Cauchy pressure.  Our calculated Cauchy pressure is positive for  Fe$_{2}$TiSn, Fe$_{2}$Ti$_{0.25}$Zr$_{0.75}$Sn$_{0.25}$Si$_{0.75}$, Fe$_{2}$Ti$_{0.25}$Zr$_{0.75}$Sn$_{0.25}$Ge$_{0.75}$,  and Fe$_{2}$Sc$_{0.25}$Ti$_{0.5}$Ta$_{0.25}$Al$_{0.5}$Bi$_{0.5}$ with the value of 12.6, 20.8, 19.2 and 12.1 GPa, respectively indicating that these systems are ductile in nature. 
The ductile or brittle behavior of a solid is estimated by B/G ratio, known as the Pugh's ratio~\cite{S.F.Pugh1954}. The critical value which separates ductile and brittle material has been evaluated to be equal to 1.75. For the ductile materials the B/G$>$1.75, and for a brittle materials, the B/G$<$1.75.  From the Table \ref{tbl:3} one can see that the value of B/G ratio for pure and isovalent/aliovalent substituted Fe$_{2}$TiSn  are larger than 1.75, meaning that these alloys are ductile in nature which is consistent with the conclusion arrived from our  Cauchy pressure analyses. Poisson's ratio provides information about the nature of chemical bonding in solids, for example, the Poisson's ratio for pure covalent crystal is 0.1 and that for the completely metallic compounds is 0.33. Our calculated Poisson's ratio for pure and isovalent/aliovalent substituted Fe$_{2}$TiSn lies in between these two values suggesting that the chemical bonding present in these systems is a mixture of covalent and metallic nature.
 
\begin{figure*}
\includegraphics[height=8cm]{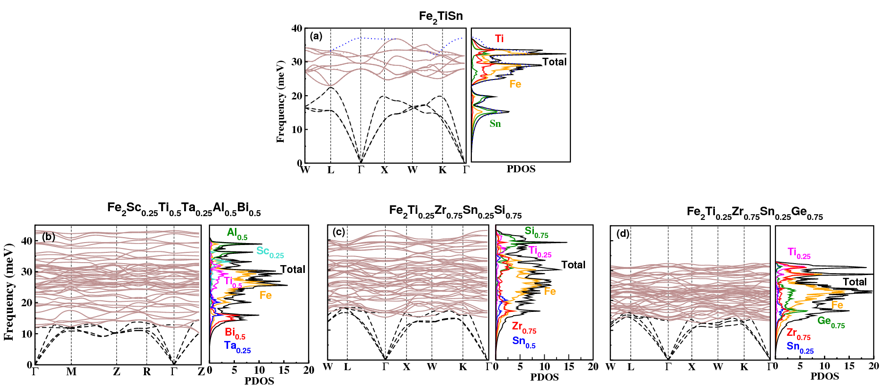}      
\caption{The calculated phonon dispersion curves and phonon density of states for (a) Fe$_{2}$TiSn, (b) Fe$_{2}$Sc$_{0.25}$Ti$_{0.5}$Ta$_{0.25}$Al$_{0.5}$Bi$_{0.5}$, (c) Fe$_{2}$Ti$_{0.25}$Zr$_{0.75}$Sn$_{0.25}$Si$_{0.75}$, and (d) Fe$_{2}$Ti$_{0.25}$Zr$_{0.75}$Sn$_{0.25}$Ge$_{0.75}$ obtained from finite difference method. The optical and acoustic modes are highlighted with brown and black color, respectively. The  non$-$analytic correction is shown by dotted blue lines.}.
\label{fgr:phn}
\end{figure*}

Figure \ref{fgr:phn} (a), (b), (c), and (d) show the phonon dispersion curves for pure Fe$_{2}$TiSn and selected isovalent/aliovalent substituted FH alloys such as , Fe$_{2}$Sc$_{0.25}$Ti$_{0.5}$Ta$_{0.25}$Al$_{0.5}$Bi$_{0.5}$, Fe$_{2}$Ti$_{0.25}$Zr$_{0.75}$Sn$_{0.25}$Si$_{0.75}$, and Fe$_{2}$Ti$_{0.25}$Zr$_{0.75}$Sn$_{0.25}$Ge$_{0.75}$.
The phonon dispersion curves are drawn along the high symmetry direction within the IBZ. The density functional perturbation theory (DFPT) with pseudopotential and plane wave methods ~\cite{refson2006variational} and also finite difference approach ~\cite{baroni2001phonons} have been used to study the phonon properties of  FH alloys. Using PBE$-$GGA functional we have calculated the phonon dispersion and phonon partial density of states (PDOS) at equilibrium lattice parameters under harmonic approximation with finite difference method as implemented in the Phonopy code. The dynamical stability of a crystal can be studied from phonon dispersion curve. The dynamically stable crystal shows all phonon frequencies in the dispersion curve positive (real), whereas the dynamically unstable crystal shows negative (imaginary) phonon frequencies in the phonon dispersion curve. All the compounds considered in the present study do not show any negative frequencies indicating that they are dynamically stable compounds at ambient condition. 
\par
Furthermore, the phonon dispersion curves can generally be divided into acoustic and optical modes. From the phonon dispersion curve of  Fe$_{2}$TiSn, one can see that the optical and acoustic modes are well separated. It is well known that the in polar solids the long range Coulomb interaction give rise to longitudinal/transverse optical splitting known as LO$-$TO splitting~\cite{baroni2001phonons} at the BZ centre. The LO$-$TO splitting is treated by including the non$-$analytical term correction in the calculation. The detailed discussion on the various approaches to estimate LO$-$TO splitting such as mixed space approach can be found in ref ~\cite{wang2016first}. A mixed$-$space approach to the DFT as implemented in the Phonopy code is used to calculate the Born effective charge and LO$-$TO splitting.
We have calculated the phonon dispersion curve for Fe$_{2}$TiSn with and without non$-$analytical term correction in our calculation.
From the Fig. \ref{fgr:phn} (a) one can see that the acoustic bands of  Fe$_{2}$TiSn are doubly degenerate along the $\Gamma-L$ direction and the maximum frequency attained by the acoustic band and optical band are around 22 meV and 37.2 meV, respectively without considering the LO$-$TO splitting into account. However taking the LO$-$TO splitting into account, one can see the splitting between longitudinal and transverse optical modes at the $\Gamma$ point. The LO$-$TO splitting frequencies at the zone centre is 5.11 meV.
\par
However, in the case of isovalent/aliovalent substituted Fe$_{2}$TiSn, we found that the substitution of a different atoms with different mass creates strong optical$-$acoustic band mixing. The optical$-$acoustic band mixing observed in the isovalent/aliovalent substituted Fe$_{2}$TiSn indicates that more phonon$-$phonon scattering is present in these systems. Hence, one can suggest that the substituted Fe$_{2}$TiSn systems will have relatively low thermal conductivity than the parent compound and this is advantageous to enhance their TE figure$-$of$-$merit.
\par
From Fig.\ref{fgr:phn} one can see that the optical$-$phonon energy is decreasing from aliovalent substituted system (Fe$_{2}$Sc$_{0.25}$Ti$_{0.5}$Ta$_{0.25}$Al$_{0.5}$Bi$_{0.5}$) to isovalent substituted system (Fe$_{2}$Ti$_{0.25}$Zr$_{0.75}$Sn$_{0.25}$Si$_{0.75}$, and Fe$_{2}$Ti$_{0.25}$Zr$_{0.75}$Sn$_{0.25}$Ge$_{0.75}$ ). This can be attributed to the fact that the equilibrium volume for isovalent system such as Fe$_{2}$Ti$_{0.25}$Zr$_{0.75}$Sn$_{0.25}$Ge$_{0.75}$ is larger than the aliovalent substituted system which causes low phonon frequencies. The small equilibrium volume of the solid is usually arising from short inter atomic distance resulting from strong chemical bonding that gives a high value of force constants and hence high phonon frequencies. 
\par
From the phonon dispersion curve we can also see that the maximum energy for acoustic mode decreases in the order from Fe$_{2}$Ti$_{0.25}$Zr$_{0.75}$Sn$_{0.25}$Si$_{0.75}$, Fe$_{2}$Ti$_{0.25}$Zr$_{0.75}$Sn$_{0.25}$Ge$_{0.75}$ and Fe$_{2}$Sc$_{0.25}$Ti$_{0.5}$Ta$_{0.25}$Al$_{0.5}$Bi$_{0.5}$, respectively. Figure \ref{fgr:phn} (a), (b), (c), and (d) (right panel) show the total and atom projected phonon density of states (PDOS)  for pure and the isovalent/aliovalet substituted Fe$_{2}$TiSn. From the total and partial PDOS we can see that the phonon modes in the acoustic region are mainly dominated by the vibrations of the heaviest atoms such as Ta, Zr, Bi, and Sn in the compounds considered in the present study. Whereas, a very small contribution can be seen from the vibration of  Ti, and Ge atoms in lower energy range. Similarly, the vibration of the phonon in mid frequency optical modes are largely owing to the vibration of the Fe atoms and small contributions of the vibration from the Ti, Zr, Bi, Si, Sn, and Ge atoms as evident from Figure \ref{fgr:phn}. Correspondingly the optical$-$phonon mode of highest energy is dominated by the vibration of the lightest atoms (Ti, Al, and Si).
\subsection{Chemical bonding analysis}
\subsubsection{Charge density, charge transfer, and electron localization function analysis}
\begin{figure*}
\centering
\includegraphics[height=5.5cm, width=12.8cm]{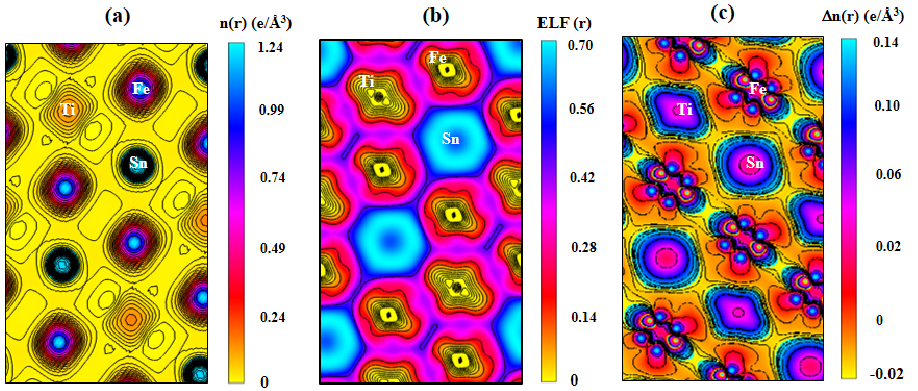}      
\caption{The calculated (a) electron density distribution, (b) electron localization function , and (c) charge transfer plot for Fe$_{2}$TiSn. The electron density, electron localization function and charge transfer plot is shown for the the plane where Fe$-$Ti and Fe$-$Sn bonds are present}.
\label{fgr:chgelf}
\end{figure*}

We turn our attention to the analysis of charge densities and related quantities such as charge transfer and electron localization function (ELF) to better understanding  the chemical bonding interactions between the constituents in Fe$_{2}$TiSn. Figure. \ref{fgr:chgelf} shows the charge density distribution of  Fe$_{2}$TiSn in appropriate plane showing the bonding interactions between the constituents. From Fig. \ref{fgr:chgelf} (a) it is apparent that the charges are largely spherically distributed at the Fe, Ti and Sn sites which is the characteristic feature for the system having ionic interactions. However, the charge density plot shows slight deviation from exact spherical distribution, which indicate finite covalent character in the Fe$-$Sn bonding. Furthermore 
our charge density plot do not help much in differentiating metallic from covalent bonding and therefore the ELF is considered as an alternative way to quantitative measure the metallicity vs covalency of a given bond.  ELF was introduced by Becke and Edgecombe ~\cite{becke1990simple} to measure the conditional probability of finding an electron in the neighborhood of another electron with the same spin. By definition, ELF is close to one in the region where electrons are paired to form covalent bond, also close to one where the unpaired lone electron of a dangling bond is localized, while it is small in low density regions. Furthermore, for homogeneous electron gas ELF is 0.5 at any density, the value close to this order indicate regions where the bonding has a metallic character. Figure. \ref{fgr:chgelf} (b)  shows the ELF for Fe$_{2}$TiSn. The maximum ELF value of around 0.70 is found in between Fe$-$Ti as well as Fe$-$Sn with a nonspherical distribution in the interstitial region is an indication for the presence of covalent character in these systems. The charge transfer plot is an another technique to analyze the bonding character in solid. The charge transfer contour is calculated by first calculating the self consistent electron density in a particular plane $\rho_{comp}$ and the electron density of the overlapping free atoms in the same plane $\rho_{atom}$  
i.e. 
$\Delta \rho(r) = \rho_{comp} - \rho_{atom}$

The charge transfer plot for Fe$_{2}$TiSn is given in Fig. \ref{fgr:chgelf} (c).  In the charge transfer plot the
 positive and negative values are associated with the charge gain and charge depletion during the formation of the solid. From  (Fig \ref{fgr:chgelf} (c)) one can see that the charge gain is mainly happening at the Ti and Sn sites ( positive charge values) while, Fe site lost charges ( negative charge value). The charge transfer distribution at Fe site is anisotropic in nature. The anisotropic charge transfer distribution between Fe$-$Ti and Fe$-$Sn indicate the presence of covalent interaction between these atoms.
In summary our calculated charge density shows that both Ti and Sn donate electrons to the Fe sites and hence we have relatively small charge at the Ti and Sn site and the transferred electrons from both Ti and Sn are accumulated at Fe site as evident from the Fig. \ref{fgr:chgelf} (a). Even though there is strong  ionic boding between Ti and Fe, there is noticeable covalent bond exists between Ti and Fe make anisotropic charge distribution at the Ti site pointing toward Fe indicating mixed iono$-$covalent character, whereas the bonding interaction between between Fe and Sn is dominantly having ionic character. The charge transfer plot given in Fig. \ref{fgr:chgelf} (c) also reflected that there is a anisotropic charge transfer distribution between Ti and Fe along with substantial charge transfer from Ti to Fe confirming the iono$-$covalent character in Fe$-$Ti bond.

\begin{table*}
\small
\centering
\caption{The calculated Bader charge (BC), Mullikan charge (MC), and Born effective charge tensor ($Z^*_{ij}$) for the constituents in Fe$_{2}$TiSn obtained from the PBE$-$GGA calculation.}
\label{tbl:4}
\begin{tabular}{c c* {3}{c c c c }}
\hline\hline\\ [-1.5ex]
Compound & Atom site & BC &MC& \multicolumn{9}{c}{$Z^*$(e)}\\
\cline{5-13} \\[-1.5ex]
 & & & &Z$_{xx}$ & Z$_{yy}$ & Z$_{zz}$ & Z$_{xy}$ & Z$_{yz}$ & Z$_{zx}$ & Z$_{xz}$ & Z$_{zy}$ & Z$_{yx}$\\
 \hline\hline\\ [-1.5ex]
  Fe$_{2}$TiSn & Fe & $-$0.529 & $-$0.730 & $-$5.055 & $-$5.055 & $-$5.055 & 0.000 & 0.000 & 0.000 & 0.000 & 0.000& 0.000 \\[2.5ex]
  & Ti & 1.047 & 1.30 & 6.084 &6.084 & 6.084 & 0.000& 0.000 & 0.000 & 0.000 & 0.000& 0.000\\[2.5ex]
 & Sn & 0.013 & 0.170& 4.032 & 4.032 &4.032 & 0.000 & 0.000 & 0.000 & 0.000 & 0.000& 0.000 \\ [2.5ex]
 \hline\\
\end{tabular}
\end{table*}

\subsubsection{Born effective charge, Bader and Mulliken effective charge analyses for Fe$_{2}$TiSn}
To reveal the bonding interaction between constituents in Fe$_{2}$TiSn we have calculated the Born effective,  Bader and Mulliken effective charge at the Fe, Ti and Sn sites and the calculated values are listed in Table~ \ref{tbl:4}. Born effective charge (BEC) denoted by Z* is a transverse or dynamical effective charge that manifests coupling between lattice displacement and electrostatic fields. Through BEC one can visualize the mixed ionic or covalent character of the bond and also investigate the lattice dynamic property of polar crystals~\cite{born1955dynamical}.
From the BEC for the atomic sites in Fe$_{2}$TiSn given in the table \ref{tbl:4} all the diagonal elements of BEC are equal and the off$-$diagonal elements are zero and this suggest that ionic interaction between these constituents is the dominant bond in this system. Especially both Ti and Sn atom donate electrons and hence the BEC is having value of  6.08 and 4.03, respectively. On the other hand, the Fe site receives electron and have the BEC value at the Fe site is large negative value of $-$5.05.
\par
Additionally the coexistence of ionic and covalent bonding in Fe$-$Ti bond can be confirmed by Bader~\cite{bader1985atoms} charge and Mulliken population analysis~\cite{mulliken1955electronic}.
Bader's atoms in molecules (AIM) charge density topology analysis  are calculated to obtain deeper insight about chemical bonding in Fe$_{2}$TiSn. The difference between the charge present in a Bader's atom and its atomic charge give the measure of the ion charge in the crystal lattice. Mullikan Charge (MC) calculated based on Mullikan population analysis for Fe$_{2}$TiSn are listed in Table \ref{tbl:4}. 

Bader charge analysis shows that Ti and Sn atom loses 1.047 and 0.013 electrons, respectively, while each Fe atom received 0.529 electrons. The calculated Mulliken effective charge shows that the charge transfer from Ti and Sn  are 1.29 and 0.17, respectively. 
From the above discussion of  BEC,  Bader effective charge and Mulliken charge analysis we qualitatively arrive at the same conclusion that both Ti and Sn donate electrons to the Fe site and also Ti is donating more electron than Sn. This again indicates the that Fe$-$Ti has strong bonding interaction than that of  Fe$-$Sn and this is consistent with the conclusion arrived from our ICOHP analysis. The electronegativity at the Fe, Ti and Sn sites are 1.83, 1.54, and 1.96, respectively which indicate that Fe is most electronegative than both Ti and Sn and hence it is expected that both Ti and Sn will donate electron to the electronegative Fe site is consistent with our BEC and Bader effective charge and Mulliken charge analysis.

\begin{figure}
\begin{center}
\includegraphics[height=7.5cm, width=7.5cm]{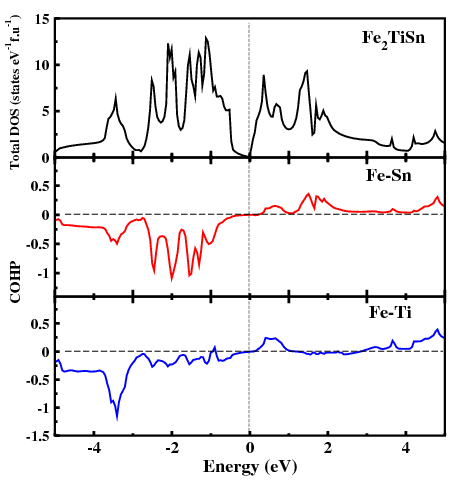}      
\caption{Projected crystal orbital Hamilton population for Fe$-$Ti and Fe$-$Sn interactions in Fe$_{2}$TiSn.  The Fermi level is set to zero}.
\label{fgr:cohpdos}
\end{center}
\end{figure}

\subsubsection{Crystal orbital Hamilton populations analysis}
Crystal orbital Hamilton populations (COHP) is the simplest way to find the bonding states between two interacting atoms in a solid ~\cite{dronskowski1993crystal}. We have calculated the COHP using Local$-$Orbitals Basis Suit Towards Electronic$–$Structure Reconstruction (LOBSTER) code using the pbeVaspFit2015 basis set ~\cite{deringer2011crystal}. COHP plot indicates the bonding, nonbonding and antibonding energy regions within a specific energy range. The negative value of COHP indicate the bonding contribution and the positive COHP value  indicate the antibonding contribution for a bonding pair. Also one can study the stability of the compound when electrons are added or removed by substitutional dopants. Figure \ref{fgr:cohpdos} shows the total density of states (DOS) and the COHP plots per bond for the nearest neighbor Fe$-$Sn and Fe$-$Ti interactions in Fe$_{2}$TiSn. The Fermi level is set to zero and  shown by the dashed vertical line. From the Fig. \ref{fgr:cohpdos} we can see that the valence band is filled with bonding states and the antibonding states are empty indicating the strong bonding interaction  for Fe$-$Sn and Fe$-$Ti bonding pairs in Fe$_{2}$TiSn. The main bonding interaction in the energy region from $-$2.5 eV to E$_{F}$ originates from  Fe$-$Sn bonds. 
\par
The bond strength between Fe$-$Sn and Fe$-$Ti bonding pairs in Fe$_{2}$TiSn are investigated by calculating the 
 integrated COHP (ICOHP) values. The ICOHP value for Fe$-$Sn and Fe$-$Ti are 2.36 and 1.98, respectively suggesting that Fe$-$Ti has strong bonding interaction than that of  Fe$-$Sn. The stronger bond strength in Fe$-$Ti bond is associated with the finite covalent bond between Fe$-$Ti bond, which is evident from the nonspherical charge density and ELF distribution at the Ti site pointed towards Fe site as evident from Fig.\ref{fgr:chgelf}.
\par
\subsection{Gr\"{u}neisen parameters and Debye Temperature from  calculated elastic properties and Phonon dispersion curve}
The anharmonic effects in the phonon spectrum due to the change in cell volume of a crystal is commonly described by Gr\"{u}neisen parameters. The first principal calculations of thermodynamic Gr\"{u}neisen parameter can be made with the quasi$-$harmonic approximation. We have employed two different approaches for calculating the  Gr\"{u}neisen parameters and Debye temperature .

\begin{figure}
 \centering
\includegraphics[height=7cm, width=9cm]{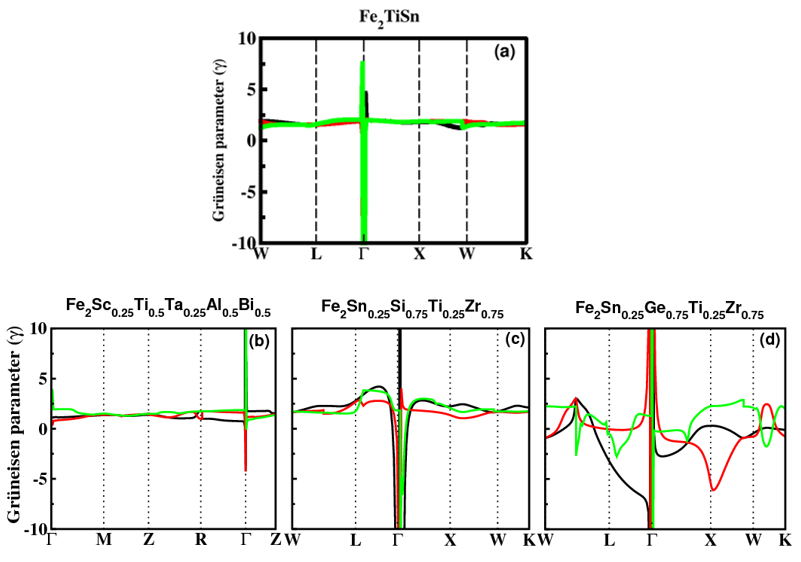}
\caption{The calculated mode Gr\"{u}neisen parameter along high symmetry directions from lattice dynamical calculations for  (a) Fe$_{2}$TiSn, (b) Fe$_{2}$Sc$_{0.25}$Ti$_{0.5}$Ta$_{0.25}$Al$_{0.5}$Bi$_{0.5}$, (c) Fe$_{2}$Ti$_{0.25}$Zr$_{0.75}$Sn$_{0.25}$Si$_{0.75}$, and (d) Fe$_{2}$Ti$_{0.25}$Zr$_{0.75}$Sn$_{0.25}$Ge$_{0.75}$. The Longitudinal acoustic (LA) and transverse acoustic (TA1 and TA2) modes are shown by Green, red and black colors, respectively.}
\label{fgr:gru}
\end{figure}
In the first approach, we have used the equation \ref{eqn25} for calculating  the Gr\"{u}neisen parameter ($\gamma$), Debye temperature ($\theta_{e}$), and acoustic Debye temperature ($\theta_{a}$)  using the calculated elastic properties such as  v$_{t}$ , v$_{l}$ and v$_{m}$  (denoted as $\theta_{e}$  and $\gamma_{e}$).  In the second method, the phonon dispersion curve calculated using lattice dynamic calculations is used to calculate the acoustic Debye temperature (denoted as $\theta_{\omega}$) and the mode Gr\"{u}neisen parameters. For calculating the mode Gr\"{u}neisen parameters, phonon calculations are performed at three different volumes i.e. one at equilibrium volume and two at slightly larger and smaller volume than the equilibrium volume, then the Gr\"{u}neisen parameters (denoted as $\gamma_{\omega}$) for each phonon mode were calculated by applying finite difference method. The calculated value of mode Gr\"{u}neisen  parameters for pure and isovalent/aliovalent substituted Fe$_{2}$TiSn obtained from both approach are listed in Table. \ref{tbl:3}.    
The mode Gr\"{u}neisen parameters for pure and isovalent/aliovalent substituted Fe$_{2}$TiSn  plotted along high$-$symmetry directions in the IBZ are shown in Fig. \ref{fgr:gru}. Longitudinal acoustic (LA) and transverse acoustic (TA1 and TA2) modes are shown by Green, red and black color, respectively.
\par
From the Table. \ref{tbl:3} one can see that the Gr\"{u}neisen  parameter calculated using elastic properties show higher value than  that calculated from the phonon dispersion curve using finite difference method for Fe$_{2}$TiSn, Fe$_{2}$Ti$_{0.25}$Zr$_{0.75}$Sn$_{0.25}$Si$_{0.75}$, and Fe$_{2}$Ti$_{0.25}$Zr$_{0.75}$Sn$_{0.25}$Ge$_{0.75}$ except for Fe$_{2}$Sc$_{0.25}$Ti$_{0.5}$Ta$_{0.25}$Al$_{0.5}$Bi$_{0.5}$ . On comparing the Gr\"{u}neisen parameters (see Table \ref{tbl:3} calculated from both the approaches discussed above one can find that the Gr\"{u}neisen parameter for isovalent/aliovalent substituted Fe$_{2}$TiSn  are generally larger than that for pure Fe$_{2}$TiSn, and the corresponding Debye temperature is lower than that of pure Fe$_{2}$TiSn. The large value of Gr\"{u}neisen parameter and lower acoustic Debye temperature suggest that isovalent/aliovalent substituted Fe$_{2}$TiSn systems will show large phonon$-$phonon scattering and hence low lattice thermal conductivity. 

\begin{figure*}
\centering
\includegraphics[height=7cm]{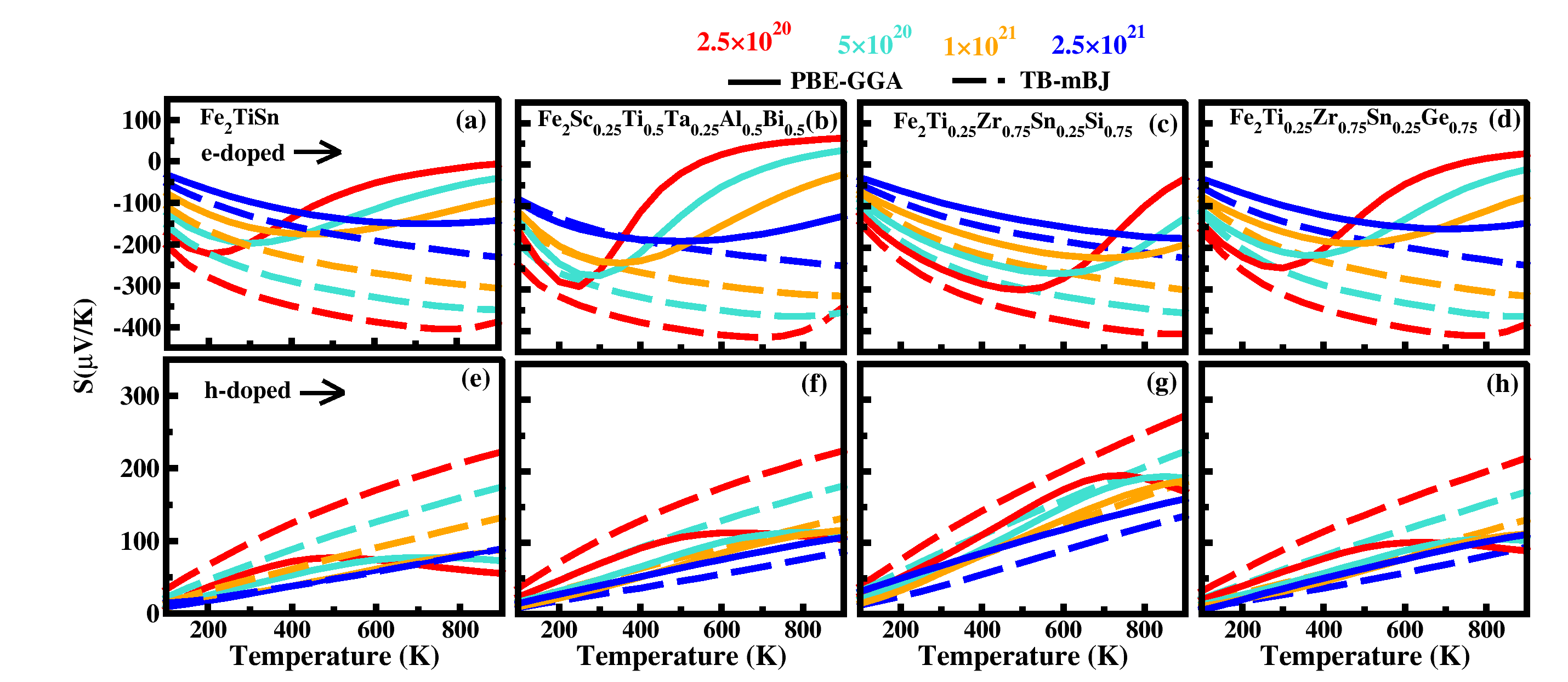}
\caption{The temperature dependence Seebeck coefficient for electron doped (upper panel) and hole doped (lower panel) Fe$_{2}$TiSn, Fe$_{2}$Sc$_{0.25}$Ti$_{0.5}$Ta$_{0.25}$Al$_{0.5}$Bi$_{0.5}$, Fe$_{2}$Ti$_{0.25}$Zr$_{0.75}$Sn$_{0.25}$Si$_{0.75}$, and Fe$_{2}$Ti$_{0.25}$Zr$_{0.75}$Sn$_{0.25}$Ge$_{0.75}$ as a function of carrier concentrations obtained from PBE$-$GGA (continuous line) and TB$-$mBJ (dashed line) calculations.}
\label{fgr:S}
\end{figure*}
\begin{figure*}
\centering
\includegraphics[height=6cm, width=13cm]{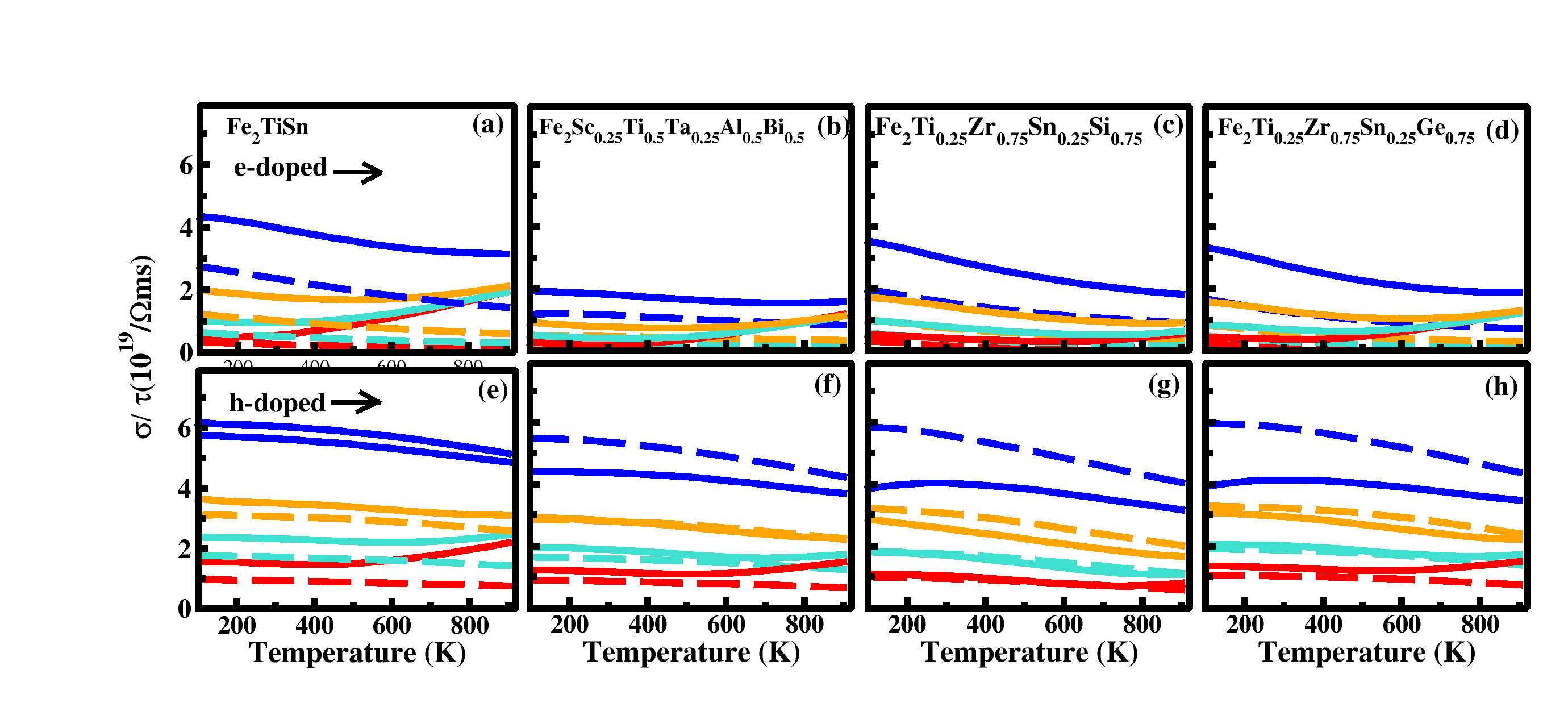}
\end{figure*}
\begin{figure*}
\centering
\includegraphics[height=6cm, width=13cm]{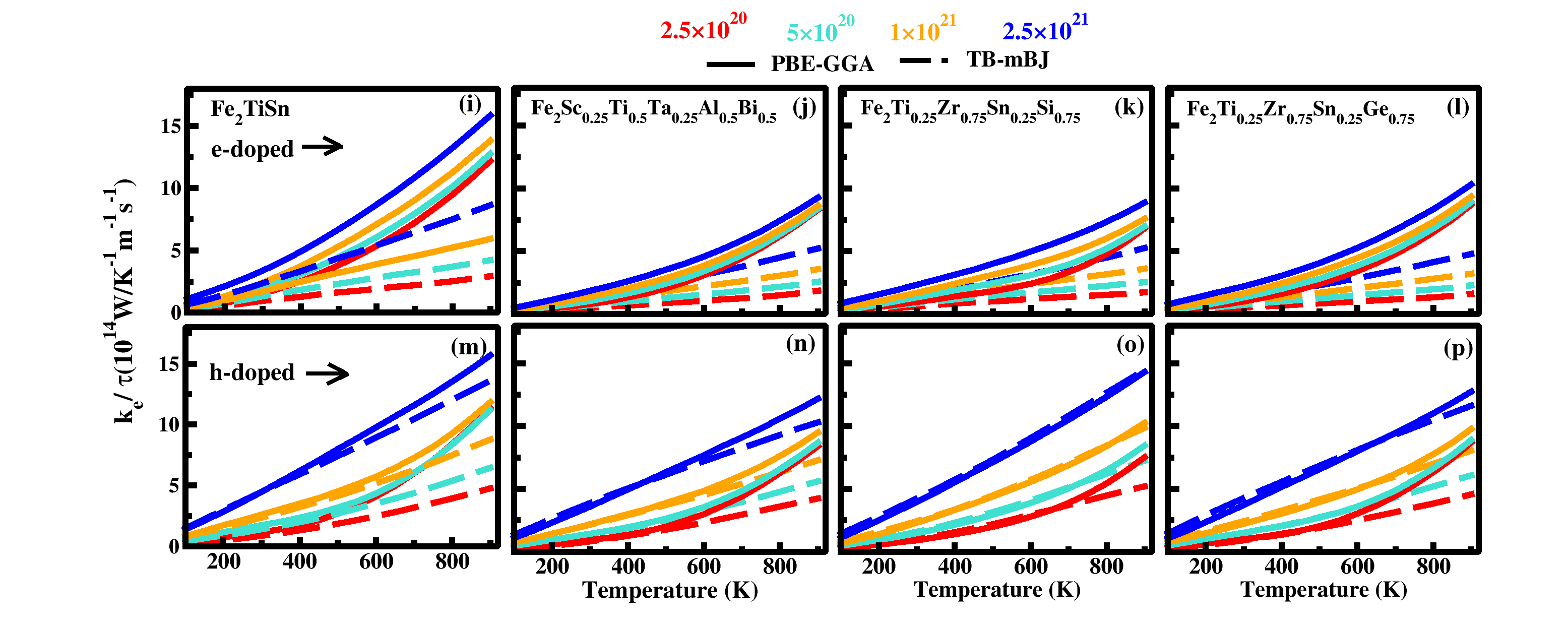}
\caption{The temperature dependent electronic part of thermal conductivity for electron doped (upper panel) and hole doped (lower panel) Fe$_{2}$TiSn, Fe$_{2}$Sc$_{0.25}$Ti$_{0.5}$Ta$_{0.25}$Al$_{0.5}$Bi$_{0.5}$, Fe$_{2}$Ti$_{0.25}$Zr$_{0.75}$Sn$_{0.25}$Si$_{0.75}$, and Fe$_{2}$Ti$_{0.25}$Zr$_{0.75}$Sn$_{0.25}$Ge$_{0.75}$ as  a function of carrier concentrations obtained from PBE$-$GGA (continuous line) and TB$-$mBJ (dashed line) calculations}
\label{fgr:rhoke}
\end{figure*}
\begin{figure*}
\centering
\includegraphics[height=8cm]{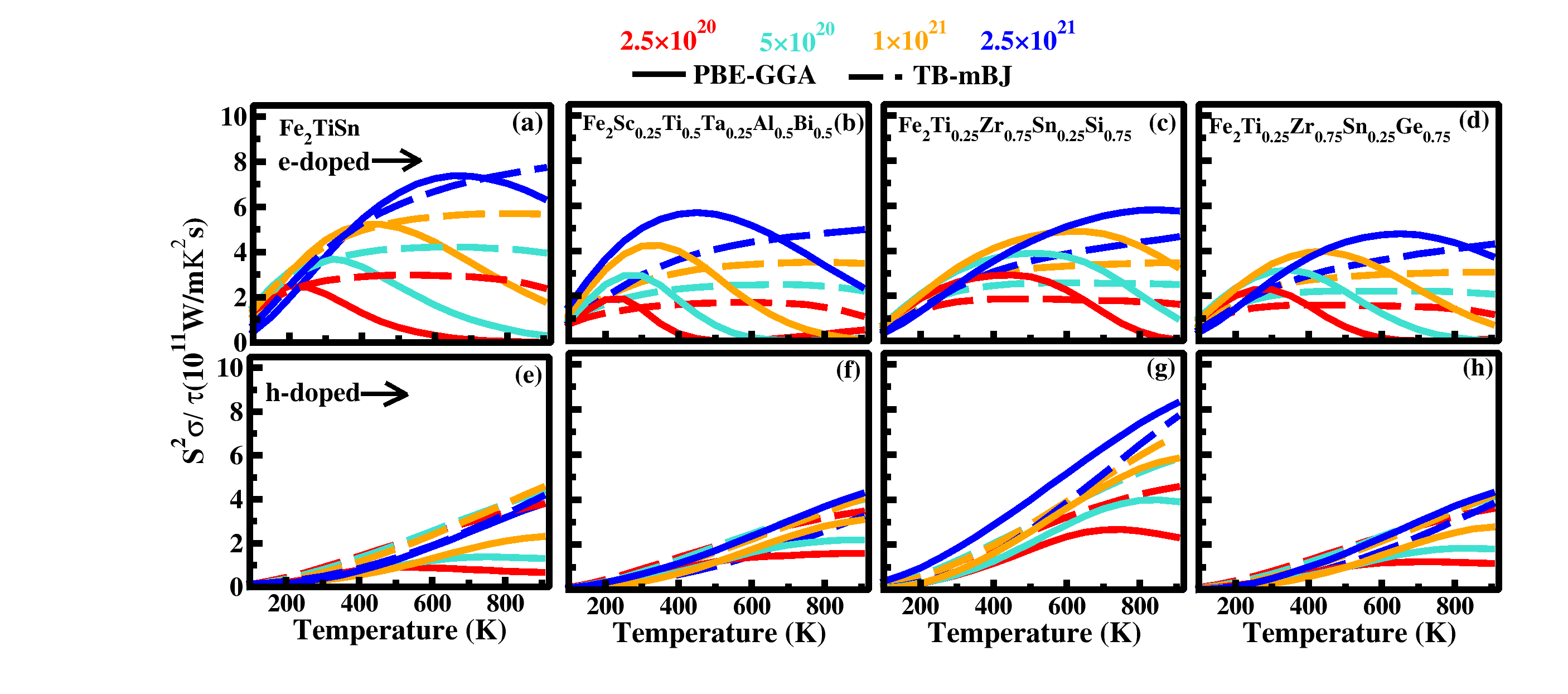}
\caption{The temperature dependent power factor for electron doped (upper panel) and hole doped (lower panel) Fe$_{2}$TiSn, Fe$_{2}$Sc$_{0.25}$Ti$_{0.5}$Ta$_{0.25}$Al$_{0.5}$Bi$_{0.5}$, Fe$_{2}$Ti$_{0.25}$Zr$_{0.75}$Sn$_{0.25}$Si$_{0.75}$, and Fe$_{2}$Ti$_{0.25}$Zr$_{0.75}$Sn$_{0.25}$Ge$_{0.75}$ as a function of carrier concentrations  obtained from PBE$-$GGA (continuous line) and TB$-$mBJ (dashed line) calculations}
\label{fgr:pf}
\end{figure*}

\subsubsection{Thermoelectric transport properties }
\label{sub:Thermoelectric transport properties}
The Seebeck coefficient and electrical conductivity for metals or degenerate semiconductors can be calculated using the following equation 
\begin{equation}\label{eqn26}
S = \frac{8\pi^2k^{2}_{B}}{3e\hbar^2}\Big(\frac{\pi}{3n}\Big)^{2/3} m^{*}_{d}T
\end{equation}
\begin{equation}\label{eqn27}
\sigma = \frac{ne^{2}\tau}{m^{*}},
\end{equation}
where  k$_{B}$, $\hbar$, e, T, n, m$^{*}_{d}$, $\sigma$, and $\tau$ are the Boltzmann constant, Planck's constant, electrical charge, absolute temperature, carrier concentration, carrier effective mass, electrical conductivity, and relaxation time of electron, respectively. The Seebeck coefficient is directly proportional to the effective mass and inversely proportional to the carrier concentration, whereas, the electronic conductivity is directly proportional to the carrier con9ocentration and inversely proportional to the effective mass. For instance, increasing electrical conductivity always results in low Seebeck coefficients and high electronic part of thermal conductivity. Therefore, to maximize the \textit{ZT} of a material, these interrelated properties such as Seebeck coefficient, electrical conductivity, and thermal conductivity need to be optimized. In this study, electronic transport properties were obtained using the band structures calculated from PBE$-$GGA and TB$-$mBJ functionals and the BoltzTraP code with a constant relaxation time of $ \tau$= 10$^{-14}$ s for the carrier concentration range from 2.5$\times$10$^{20}$ to 2.5$\times$10$^{21}$ cm$^{-3}$.
Figure. \ref{fgr:S} shows the calculated Seebeck coefficient as a function of temperature for several carrier concentrations for h$-$doped and e$-$doped conditions in Fe$_{2}$TiSn, Fe$_{2}$Sc$_{0.25}$Ti$_{0.5}$Ta$_{0.25}$Al$_{0.5}$Bi$_{0.5}$, Fe$_{2}$Ti$_{0.25}$Zr$_{0.75}$Sn$_{0.25}$Si$_{0.75}$, and Fe$_{2}$Ti$_{0.25}$Zr$_{0.75}$Sn$_{0.25}$Ge$_{0.75}$ using the electronic structure obtained from PBE$-$GGA and TB$-$mBJ exchange$-$correlation potential with carrier concentration between 2.5$\times$10$^{20}$ to 2.5$\times$10$^{21}$ cm$^{-3}$. It may be noted that the calculated Seebeck coefficient using PBE$-$GGA and TB$-$mBJ functionals in this figure are showing completely different temperature dependent behavior. 
 \par
To find the optimum carrier concentrations to obtain maximum \textit{ZT}, it is essential to investigate how the TE properties change with electron/hole doping. The negative and positive values of  the Seebeck coefficients suggest their n$-$type and p$-$type conducting character. From Fig. \ref{fgr:S}, we can see that the Seebeck coefficients decrease with increasing electron/hole concentration irrespective of the exchange correlation potential we have used to estimate the electronic structure and hence the transport properties. In  e$-$doped conditions, the Seebeck coefficient calculated from PBE$-$GGA first increases with temperature and then decreases at high temperatures. The peak value in the S(T) curve calculated from PBE$-$GGA exists at a lower temperature at low carrier concentration (2.5$\times$10$^{20}$ cm$^{-3}$)  for Fe$_{2}$TiSn, Fe$_{2}$Sc$_{0.25}$Ti$_{0.5}$Ta$_{0.25}$Al$_{0.5}$Bi$_{0.5}$, Fe$_{2}$Ti$_{0.25}$Zr$_{0.75}$Sn$_{0.25}$Si$_{0.75}$, and Fe$_{2}$Ti$_{0.25}$Zr$_{0.75}$Sn$_{0.25}$Ge$_{0.75}$ and the corresponding peak values (at temperature) in these compounds are 228 (at 200\,K), 298 (at 250\,K), 300 (at 537\,K) and 253 V/K (at 300\,K), respectively. In e$-$doped conditions, the Seebeck coefficient calculated based on TB$-$mBJ potential increases sharply with temperature as compared to that using PBE$-$GGA potential for both pure and the isovalent/aliovalent substituted Fe$_{2}$TiSn as evident from Figs.\,\ref{fgr:S}(a)$-$(d).
 \par
In  h$-$doped condition, the Seebeck coefficients calculated using PBE$-$GGA functional for Fe$_{2}$TiSn,Fe$_{2}$Sc$_{0.25}$Ti$_{0.5}$Ta$_{0.25}$Al$_{0.5}$Bi$_{0.5}$, Fe$_{2}$Ti$_{0.25}$Zr$_{0.75}$Sn$_{0.25}$Si$_{0.75}$, and Fe$_{2}$Ti$_{0.25}$Zr$_{0.75}$Sn$_{0.25}$Ge$_{0.75}$ are sharply increasing with increase of temperature and reaches maximum value (at temperature) respectively of 80 (at 500\,K), 100 (at 600\,K), 200 (at 750\,K) and 95 V/K  (at 600\,K) at low carrier concentration of 2.5$\times$10$^{20}$ cm$^{-3}$ and then slowly decreases beyond this temperature. It is worthy to note that, in both e$-$doped and h$-$doped conditions, the Seebeck coefficient calculated using TB$-$mBJ potential is larger than that using PBE$-$GGA functional. The larger band gap value and relatively flat band at the CB edges obtained using TB$-$mBJ are responsible for such high Seebeck coefficient. From the calculated Seebeck coefficient as a function of temperature and carrier doping we have found that the high value of Seebeck coefficient is obtained when these systems are doped with electrons.
\par
Figure. \ref{fgr:rhoke} shows the electrical conductivity (a)$-$(h) and electronic thermal conductivity divided by relaxation time $\tau$ (i)$-$(p) for Fe$_{2}$TiSn, Fe$_{2}$Sc$_{0.25}$Ti$_{0.5}$Ta$_{0.25}$Al$_{0.5}$Bi$_{0.5}$, Fe$_{2}$Ti$_{0.25}$Zr$_{0.75}$Sn$_{0.25}$Si$_{0.75}$, and Fe$_{2}$Ti$_{0.25}$Zr$_{0.75}$Sn$_{0.25}$Ge$_{0.75}$  as a function of temperature with carrier concentrations varying between 2.5$\times$10$^{20}$ cm$^{-3}$ to 2.5$\times$10$^{21}$  cm$^{-3}$ calculated using both PBE$-$GGA (continuous line) and TB$-$mBJ functionals (dashed line). From  Fig.\, \ref{fgr:rhoke} (a$-$h) we can see that the carrier concentration dependent electrical conductivity shows an opposite trend compared to that of Seebeck coefficient (see Fig.\,\ref{fgr:S}(a)$-$(d))  i.e., the electrical conductivity values increase with increase of carrier concentration. In e$-$doped condition, the electrical conductivity calculated with the PBE$-$GGA potential are slightly higher value than those obtained using the TB$-$mBJ functional for both pure and isovalent/aliovalent substituted Fe$_{2}$TiSn.
From Fig. \ref{fgr:rhoke} (a$-$h) we can see that, in  hole doped condition, the electrical conductivity calculated using PBE$-$GGA and TB$-$mBJ show a very small difference for Fe$_{2}$TiSn except at the carrier concentration 2.5$\times$10$^{20}$  cm$^{-3}$. Similarly, in hole doped condition in Fig. \ref{fgr:rhoke} (i$-$p), in the case of isovalent/aliovalent substituted systems, the electrical conductivity calculated from both the functional show a very small difference except at the carrier concentration of 2.5$\times$10$^{21}$  cm$^{-3}$ where it shows a large difference. The electrical conductivity calculated using PBE$-$GGA and TB$-$mBJ functionals for pure and isovalent/aliovalent substituted Fe$_{2}$TiSn show a higher value at high hole concentration (2.5$\times$10$^{21}$  cm$^{-3}$).
\par
The electrical conductivities for pure and substituted systems are found to be high value for h$-$doped condition than the e$-$doped condition irrespective of the exchange correlation functionals we have used to evaluate them and this is due to the fact that the flat bands near the CB edge in all these systems results in high electron effective mass value than the hole effective mass and hence the $\sigma$ calculated  in the e$-$doped system are lower than those in the h$-$doped condition.
Figure \ref{fgr:rhoke} (i$-$p) shows the electronic part of  thermal conductivity $\kappa_{e}$ as a function of temperature obtained using  PBE$-$GGA and TB$-$mBJ functional for pure and isovalent/aliovalent substituted Fe$_{2}$TiSn. The variation in the value $\kappa_{e}$ with temperature is more or less the same for both h$-$doped and e$-$doped conditions for all these systems irrespective of the exchange$-$correlation functional we have used to evaluate them. The over all trend in the temperature dependence of  $\kappa_{e}$ is that it increases with temperature almost linearly and reaches maximum value at high temperature in all the compositions considered in the present. In the e$-$doped conditions, the calculated $\kappa_{e}$ for pure and substituted Fe$_{2}$TiSn show higher value when we obtain it using PBE$-$GGA than using TB$-$mBJ functional. However, in h$-$doped condition, we have found very small difference between the  $\kappa_{e}$ value calculated by using both the functional except at low carrier concentration ~2.5$\times$10$^{20}$  cm$^{-3}$ (see Figure  \ref{fgr:rhoke} (i$-$p)).
\par
Figure \ref{fgr:pf} (a)$-$(h) show the power factor (PF) calculated with PBE$-$GGA (continuous line)  and TB$-$mBJ (dashed line) functionals for pure and isovalent/aliovalent substituted Fe$_{2}$TiSn as a function of temperature for both electron doped and hole doped conditions. It is clear from these plots that in e$-$doped condition, PF calculated from PBE$-$GGA functional increases with temperature and reaches a maximum value at a certain temperature and then slowly decreases at a higher temperature for carrier concentration 2.5$\times$10$^{20}$  cm$^{-3}$ to 2.5$\times$10$^{21}$ cm$^{-3}$ in all the systems considered in the present study. Also, one can notice that the PF peak shift to a high temperature range with increasing carrier concentration. For the electron concentration between 2.5$\times$10$^{20}$ cm$^{-3}$ to 1$\times$10$^{21}$ cm$^{-3}$ the calculated PF using TB$-$mBJ increases with temperature up to 300\,K and remain almost constant at high temperatures for pure as well as isovalent/aliovalent substituted Fe$_{2}$TiSn. However, for the electron concentration 2.5$\times$10$^{21}$ cm$^{-3}$ the PF value increase with temperature continuously (see Figure  \ref{fgr:rhoke} (i$-$p)).
\par
In h$-$doped conditions the PF calculated with TB$-$mBJ functional increases with temperature and reached a maximum value at high temperature. However, PF calculated with PBE$-$GGA functional increases with temperature and then slowly decreases at high temperature except for the concentration 2.5$\times$10$^{21}$ where the PF increases continuously with temperature. The maximum value in the temperature dependent PF curve depends on the value of Seebeck coefficient as well as the electrical conductivity at that condition. For pure Fe$_{2}$TiSn, as per the calculation based on TB$-$mBJ functional, the maximum value of PF in both e$-$doped and h$-$doped conditions are 7.86 and 4.7 10$^{11}$W/mK$^{2}$s, respectively at high carrier concentration (at 2.5$\times$10$^{20}$  cm$^{-3}$). However, in the case of  Fe$_{2}$Sc$_{0.25}$Ti$_{0.5}$Ta$_{0.25}$Al$_{0.5}$Bi$_{0.5}$, Fe$_{2}$Ti$_{0.25}$Zr$_{0.75}$Sn$_{0.25}$Si$_{0.75}$, and Fe$_{2}$Ti$_{0.25}$Zr$_{0.75}$Sn$_{0.25}$Ge$_{0.75}$, as per our calculations based on PBE$-$GGA functional, the maximum PF in e$-$doped (h$-$doped) conditions are 5.94 (4.5), 6.02(8.4), and  4.98(4.5) 10$^{11}$W/mK$^{2}$s, respectively at the carrier concentration 2.5$\times$10$^{21}$ cm$^{-3}$. From the above observations it is evident that the highest PF is found in broad temperature range when we dope our systems with  high carrier concentration i.e. around 10$^{21}$  cm$^{-3}$ . Also, it should be noted that, high value of  PF does not guarantee a high \textit{ZT} value. One would expect high \textit{ZT} in a material when it possess low lattice thermal conductivity apart from large power factor. So, we explore the lattice part of thermal conductivity in these systems in the following section.

\begin{figure}[h]
\centering
\includegraphics[height=8cm]{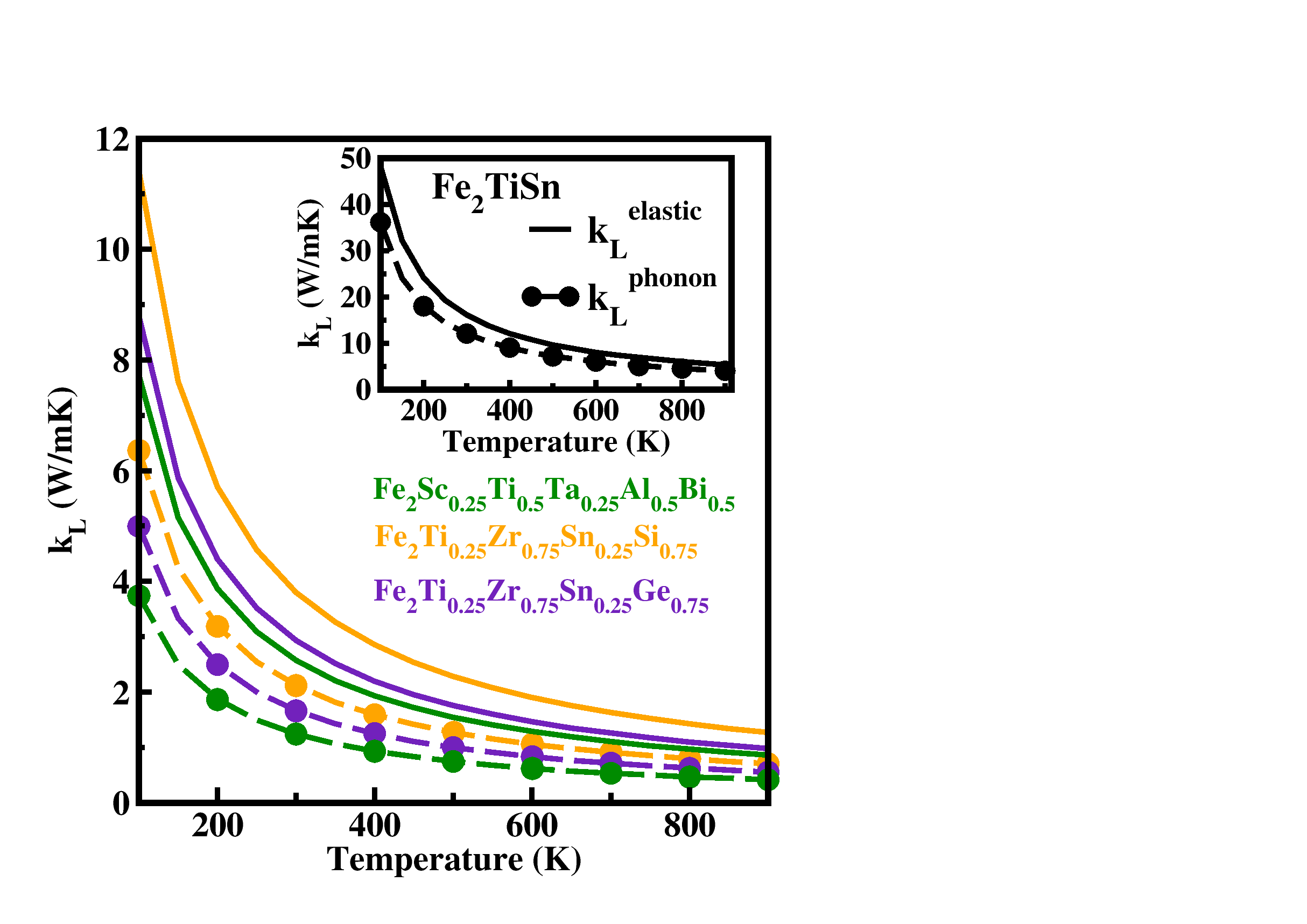}
\caption{The comparison of calculated lattice part of thermal conductivity ($\kappa_{L}$) obtained based on Slack's equation  using elastic constant data ( $\kappa_{L}^{elastic}$; continuous line) and phonon dispersion curves ($\kappa_{L}^{phonon}$; dashed line) for pure Fe$_{2}$TiSn and isovalent/aliovalent substituted Fe$_{2}$TiSn as a function of temperature obtained from PBE$-$GGA method.}
\label{fgr:kl}
\end{figure}

\begin{figure*}
 \centering
\includegraphics[height=4.5cm, width=15cm]{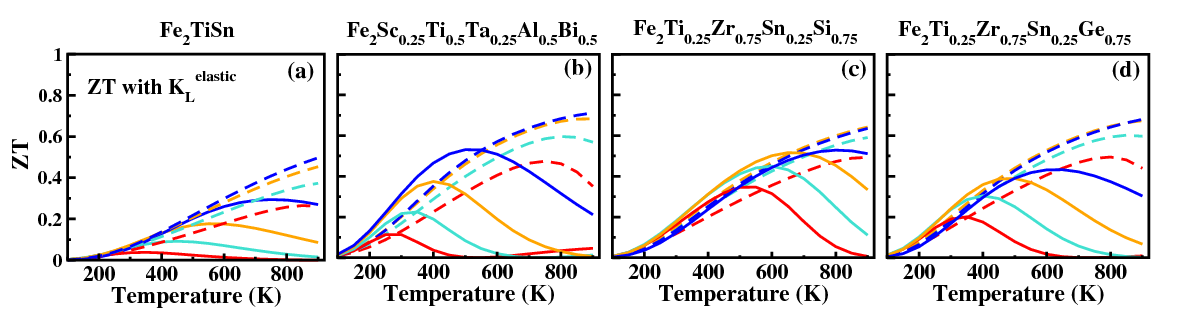}
\end{figure*}
\begin{figure*}
 \centering
\includegraphics[height=4.7cm, width=15cm]{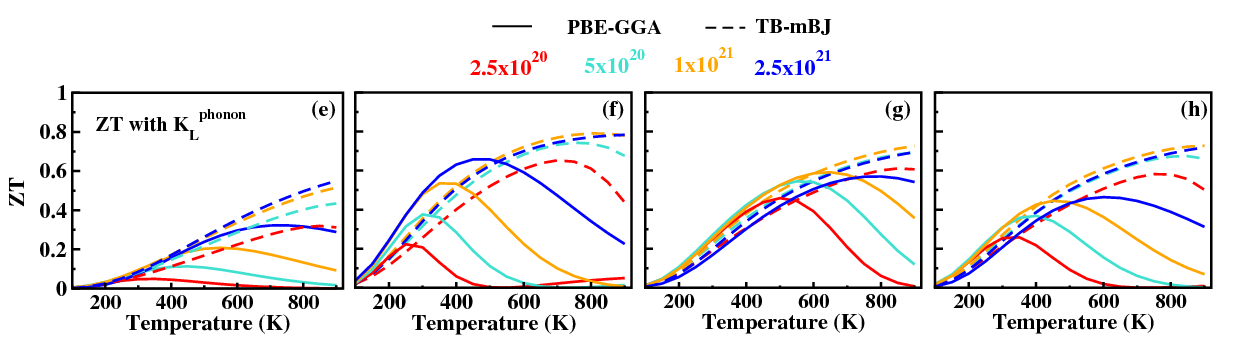}
\caption{The comparison of calculated TE figure$-$of$-$merit \textit{ZT} using lattice part of thermal conductivity ($\kappa_{L}$) obtained based on Slack's equation  using elastic constant data ( $\kappa_{L}^{elastic}$; upper panel i.e (a)$-$(d)) and  phonon dispersion curves ($\kappa_{L}^{phonon}$; lower panel i.e. (e)$-$(h)) for pure Fe$_{2}$TiSn and isovalent/aliovalent substituted Fe$_{2}$TiSn as as a function of carrier concentrations under e$-$doped and h$-$doped conditions obtained from PBE$-$GGA (continuous line) and TB$-$mBJ (dashed line) calculations, respectively.}
\label{fgr:ZT}
\end{figure*}

\subsection{Lattice thermal conductivity and thermoelectric figure$-$of$-$merit}
Materials with low thermal conductivity are of great interest for higher efficiency thermoelectrics. The variation of $\kappa_{e}$ with temperature as a function of carrier concentation for pure and isovalent/aliovalent substituted Fe$_{2}$TiSn with the considered two different exchange$-$correlation functionals are discussed in the previous section. Materials with high atomic mass, weak interatomic bonding, complex crystal structure and high anharmonicity generally have low  $\kappa_{L}$  value. Generally, for crystalline materials, lattice thermal conductivity decreases inversely with the temperature at low temperature and incontrast, at high temperatures it shows constant or very weak temperature dependence, which resembles glass$-$like behavior. The lattice part of thermal conductivity can be calculated by performing the full iterative solution to the phonon Boltzmann transport equation (BTE) with the help of open$-$source code such as Phono3py~\cite{togo2015distributions}, ShengBTE~\cite{li2014shengbte}, almaBTE~\cite{tadano2014anharmonic}, and PhonTS~\cite{chernatynskiy2015phonon}. However, solving the BTE is computationally expensive. In 1973, Slack~\cite{slack1973nonmetallic} has provided a simplest approach for calculating the $\kappa_{L}$  which is given as
 \begin{equation} \label{eqn28}
k_{L}=A\dfrac{\bar{M_{a}}\delta\theta\textsubscript{a}^{3}}{\gamma^{2}Tn^{2/3}},
\end{equation}
where $\bar{M_{a}}$, $\theta_{a}$, $\delta$, $n$, and $\gamma$ are average mass per atom in the crystal,
acoustic Debye temperature, cube root of the average volume per atom,  number of atoms in the primitive unit cell and Gr\"{u}neisen parameter, respectively, and $A$ is a physical quantity which can be calculated as A$=\dfrac{2.43\times10^{-8}}{1-0.514/\gamma+0.228/\gamma^{2}}$
\par
Figure \ref{fgr:kl} shows the theoretically calculated lattice thermal conductivity by inserting the Debye temperature and Gr\"{u}neisen parameter estimated from first$-$principles method using two different approaches, one from the calculated single crystal elastic constants c$_{ij}$ and the other from quasi$-$harmonic phonon calculations with volume into Slack's formula \ref{eqn28}.  
In the first method we have used the calculated elastic constants (the bulk and shear moduli) to calculate the Debye temperature and Gr\"{u}neisen parameter. It is to be noted that the Slack's model only considers acoustic contributions. Therefore in the second method, we have calculated the Debye temperature from the phonon dispersion curve by taking only the highest frequency of the acoustic mode using $\theta_{D}=\omega_{D}/k_{B}$ (where $\omega_{D}$ is the maximum value of acoustic frequency obtained from the phonon dispersion curve) and the Gr\"{u}neisen parameter is calculated from the quasi$-$harmonic phonon calculations through the phonon dispersion curves obtained at different volumes. The variation in $\kappa_{L}$ with temperature obtained from above two approaches for pure and the isovalent/aliovalent substituted Fe$_{2}$TiSn are much alike over the entire temperature range. The calculated $\kappa_{L}$ value for Fe$_{2}$TiSn calculated from Slack's approach with input from elastic constants calculations (phonon dispersion relation) are 8.04 (6.01) and 5.36 (4.01) W/mK at 600\,K  and 900\,K, respectively.
\par
However, we have observed that at 300 \,K, our calculated $\kappa_{L}$ value using elastic constants (phonon dispersion curve)  is around 14.8 (11.6 ) W/mK, and is much  higher value than the corresponding experimentally reported value of around 7$-$8 W/mK ~\cite{voronin2017preparation}. It may be noted that the experimentally prepared samples will involve imperfections and various defects those will reduce the measured  $\kappa_{L}$  value and this could explain the large difference between the experimentally measured and our theoretically calculated  $\kappa_{L}$  values. However, our calculated  $\kappa_{L}$  value will be applicable to compare with that of defect free single crystal. From  Fig. \ref{fgr:kl} one can see that the $\kappa_{L}$ estimated using both methods the above mentioned two approaches show a large difference at low temperatures. Especially the  $\kappa_{L}$  obtained based on the input from single crystal elastic constant calculation is always higher than that obtained based on calculated phonon dispersion curve. However, at higher temperatures, this difference gets reduced. Similarly, in the case of isovalent/aliovalent substituted Fe$_{2}$TiSn, the  calculated $\kappa_{L}$ value are small at higher temperatures and also their temperature dependent variation is very small at high temperatures. Compared to pure Fe$_{2}$TiSn, the isovalent/aliovalent substituted Fe$_{2}$TiSn shows smaller $\kappa_{L}$ values irrespective of the computational approach we have used to estimate the same. For instant, in the case of Fe$_{2}$Sc$_{0.25}$Ti$_{0.5}$Ta$_{0.25}$Al$_{0.5}$Bi$_{0.5}$,  Fe$_{2}$Ti$_{0.25}$Zr$_{0.75}$Sn$_{0.25}$Si$_{0.75}$, and Fe$_{2}$Ti$_{0.25}$Zr$_{0.75}$Sn$_{0.25}$Ge$_{0.75}$ the calculated  $\kappa_{L}$ using $\kappa^{elastic}_{L}$ ($\kappa^{phonon}_{L}$) are 1.47 (0.62), 1.90 (1.06) and 1.29 (0.83) W/mK at 600\,K and 0.98 (0.42, 1.27 (0.71) and 0.86 (0.56) W/mK at 900\,K, respectively. The lower $\kappa_{L}$ value for isovalent/aliovalent substituted Fe$_{2}$TiSn compared with pure system is due to lower acoustic Debye temperature and high Gr\"{u}neisen parameter (see Table \ref{tbl:3}).
\par
Figure \ref{fgr:ZT} shows the temperature dependence of the TE figure$-$of$-$merit calculated by using PBE$-$GGA and TB$-$mBJ functional at various carrier concentrations for pure and isovalent/aliovalent substituted Fe$_{2}$TiSn. The TE figure$-$of$-$merit is calculated by substituting S, $\sigma$, T, $\kappa_{e}$, and the $\kappa_{L}$ into the equation \ref{eqn1} dealing with \textit{ZT}. Here we have evaluated the \textit{ZT} value by including the $\kappa_{L}$ calculated using two methods, one based on calculated elastic properties ($\kappa^{elastic}_{L}$) and the other by using phonon dispersion curve ($\kappa^{phonon}_{L}$) and the details of these calculations are given in sec.\ref{sub:Mechanical stability and lattice dynamic calculation for pure and isovalent/aliovalent substituted Fe$_{2}$TiSn}.

Let us first discuss the calculated \textit{ZT} using $\kappa^{elastic}_{L}$ for both pure and isovalent/aliovalent substituted Fe$_{2}$TiSn with the electronic part obtained using both PBE$-$GGA and TB$-$mBJ functionals.
From the Fig. \ref{fgr:ZT} (a$-$d) one can see that the temperature dependent \textit{ZT} curve obtained using $\kappa^{elastic}_{L}$ with PBE$-$GGA for Fe$_{2}$TiSn and the isovalent/aliovalent substituted Fe$_{2}$TiSn systems reach the maximum value at low temperature for low carrier concentration and the peak shifts gradually to the high temperature range with increasing carrier concentration. Similarly, the \textit{ZT} calculated for all these systems using TB$-$mBJ functional for the calculation of electronic part and PBE$-$GGA for calculating lattice part also follow the same trend that the peak in the  \textit{ZT} systematically shifted to higher temperature with carrier concentration. A similar trend has also been observed for the \textit{ZT} calculated using $\kappa^{phonon}_{L}$. However, the \textit{ZT} values calculated using $\kappa^{phonon}_{L}$ for all these systems with various carrier concentration show larger value than those using $\kappa^{elastic}_{L}$ irrespective of the exchange$-$correlation functional used to estimate the electronic part part of thermal conductivity.
\par
The maximum \textit{ZT} value (at temperatures ) for Fe$_{2}$TiSn, Fe$_{2}$Sc$_{0.25}$Ti$_{0.5}$Ta$_{0.25}$Al$_{0.5}$Bi$_{0.5}$, Fe$_{2}$Ti$_{0.25}$Zr$_{0.75}$Sn$_{0.25}$Si$_{0.75}$ and Fe$_{2}$Ti$_{0.25}$Zr$_{0.75}$Sn$_{0.25}$Ge$_{0.75}$ using $\kappa^{elastic}_{L}$ with PBE$-$GGA functional are 0.31 (at 700\,K), 0.54 (at 500\,K), 0.53 (at 800\,K), and 0.44 (at 600\,K), respectively, and that with TB$-$mBJ functional are 0.50, 0.71, 0.65 and 0.69 at 900\,K, respectively, for the carrier concentration (2.5)$\times$10$^{21}$ cm$^{-3}$. Similarly, the peak in the \textit{ZT} curve with temperature for Fe$_{2}$TiSn,  Fe$_{2}$Sc$_{0.25}$Ti$_{0.5}$Ta$_{0.25}$Al$_{0.5}$Bi$_{0.5}$, Fe$_{2}$Ti$_{0.25}$Zr$_{0.75}$Sn$_{0.25}$Si$_{0.75}$ and Fe$_{2}$Ti$_{0.25}$Zr$_{0.75}$Sn$_{0.25}$Ge$_{0.75}$  using $\kappa^{phonon}_{L}$ using  PBE$-$GGA functional are 0.35 (at 750\,K), 0.68 (at 460\,K), 0.60 (at 600\,K), and 0.47 (at 600\,K),  respectively and that with TB$-$mBJ functional are 0.55, 0.80, 0.74 and 0.75 at 900\,K, respectively for the carrier concentration (2.5)$\times$10$^{21}$ cm$^{-3}$. 
Among  the isovalent/aliovalent substituted Fe$_{2}$TiSn, the aliovalent substituted system (Fe$_{2}$Sc$_{0.25}$Ti$_{0.5}$Ta$_{0.25}$Al$_{0.5}$Bi$_{0.5}$) shows the high \textit{ZT} value of  0.81 at 900\,K using  ($\kappa^{phonon}_{L}$) with TB$-$mBJ functional for evaluating electronic part of thermal conductivity, PF and $\kappa_{e}$. 
 \par
From the calculated \textit{ZT} values of isovalent/aliovalent substituted Fe$_{2}$TiSn we found that the substitution enhance the\textit{ ZT} value compared with that of Fe$_{2}$TiSn. However, within the isovalent substituted systems, the calculated \textit{ZT} value is almost the same irrespective of the substituents used in the calculation. But, our calculations show that the aliovalent substituted systems  show higher \textit{ZT} value than that of parent and isovalent substituted Fe$_{2}$TiSn. It is worth to mention that the aliovalent substituted Fe$_{2}$TiSn systems such as Fe$_{2}$Sc$_{0.25}$Ti$_{0.5}$Ta$_{0.25}$Al$_{0.5}$Bi$_{0.5}$  shows substantial  improvement of \textit{ZT} as compared to the pure as well as isovalent substituted Fe$_{2}$TiSn.  This could be explained due to the fact that the calculated lattice thermal conductivity for aliovalent substituted Fe$_{2}$TiSn is lower than that of pure and isovalent substituted Fe$_{2}$TiSn. The main reason for the reduction in lattice thermal conductivity is owing to the fact that the Bi and Ta atoms are heavier than the Si/Ge atoms which creates more phonon$-$phonon scattering center and hence reducing the lattice thermal conductivity. Due to the fluctuation in the atomic mass, a much lower $\kappa_{L}$ value is observed in the substituted systems compared to that for pure Fe$_{2}$TiSn, which makes the substituted systems most promising candidates for high efficient TE materials.
Hence, we conclude that high \textit{ZT} in aliovalent/isovalent systems is possible at high electron carrier concentrations. The overall observation shows that the \textit{ZT} calculated with the TB$-$mBJ potential is higher than those calculated with the PBE$-$GGA functional.
\section{Conclusion}
The present study we have theoretically investigated the multinary substituted full Heusler alloy Fe$_{2}$TiSn  with the isovalent/aliovalent substitution preserving the 24 valence electron count rule. The electronic structure and transport properties of pure and isovalent/aliovalent substituted Fe$_{2}$TiSn were studied in detail by including lattice part of thermal conductivity explicitly in to the calculation with results from single crystal elastic constant calculation and phonon dispersion curve. The band structure using PBE$-$GGA and TB$-$mBJ functional shows the semiconducting behavior for pure and  isovalent/aliovalent substituted Fe$_{2}$TiSn fulfilling 24 VEC rule. The calculated band gap values using PBE$-$GGA and TB$-$mBJ functional are in the range of  0.02$-$0.22 and 0.41$-$0.74 e\,V respectively for these systems. 
The isovalent/aliovalent substitution at the Ti and Sn site of Fe$_{2}$TiSn create degenerate flat bands in the vicinity of the conduction band edge which enhanced the power factor and and along with reduction in lattice part of thermal conductivity by substitution  showed excellent thermoelectric transport properties in n$-$type doping condition. 
\par
The  carrier effective mass values for pure as well as isovalent substituted Fe$_{2}$TiSn are calculated by fitting the calculated band extrema of the band structure obtained using the PBE$-$GGA and TB$-$mBJ functional within the parabolic approximation. The carrier effective mass values calculated from the band structure obtained using TB$-$mBJ functional shows higher value than those obtained using PBE$-$GGA functional for all the compounds considered in the present study. The calculated elastic stiffness constant c$_{ij}$  confirmed that the investigated compounds are mechanically stable. Bulk modulus, shear modulus, Young's modulus and Poisson's ratio were calculated from the calculated elastic stiffness modulii using PBE$-$GGA functional. The calculated phonon dispersion curves and phonon density of states using finite difference method with supercell approach show no negative frequency in all the compounds considered in this study and hence confirmed their dynamical stability. The chemical bonding character of  pure Fe$_{2}$TiSn is iono$-$covalent according to analyses of charge density distribution, electron localization function, charge transfer, Mulliken population, Bader charge and Born effective charge. 
The TE transport properties such as Seebeck coefficient, electrical conductivity, power factor are obtained from the band structure obtained using both BE$-$GGA with the TB$-$mBJ functionals as a function of temperature and carrier density and are discussed.  In addition, the comparison of the calculated TE  transport properties obtained using the calculated thermal conductivity based on elastic constants as well as phonon dispersion curve with electronic part calculated from PBE$-$GGA and  TB$-$mBJ functionals showed a significant increase of the \textit{ZT} values when we use the TB$-$mBJ functional. Among all the systems considered in the present study Fe$_{2}$Sc$_{0.25}$Ti$_{0.5}$Ta$_{0.25}$Al$_{0.5}$Bi$_{0.5}$   shows the high \textit{ZT} value of 0.81 at 900\,K  when we use lattice part of thermal conductivity along with electronic contribution obtained using TB$-$mBJ functional. From detailed analysis we demonstrates that the isovalent/aliovalent substitutions is one key approach reducing the lattice part of thermal conductivity by mass fluctuation and maximizing the scattering centre in thermoelectric materials and hence increase the \textit{ZT} value. The present investigation of structural, electronic, lattice dynamic and thermoelectric transport properties of isovalent/aliovalent substituted Fe$_{2}$TiSn will be useful in designing future TE materials based on 24 VEC full$-$Heusler alloys with higher efficiency and motivate experimentalists to synthesis such compounds.
\section*{Acknowledgment}
The authors are grateful to the Science and Engineering Research Board (SERB) a stationary body of  Department of Science and Technology, Government of India, for the funding support  under the scheme SERB$-$Overseas Visiting Doctoral Fellowship(OVDF) via award  no. ODF/2018/000845  and the Research Council of Norway for providing the computer time (under the project number NN2875k) at the Norwegian supercomputer facility. The authors would also like to acknowledge the SERB$-$Core Research Grant (CRG) vide file no.CRG/2020/001399.



\bibliography{Arxiv}
\end{document}